\documentclass[twocolumn,english,aps,superscriptaddress,floatfix,showpacs,prb,final]{revtex4}
\usepackage{mathptmx}

\usepackage[T1]{fontenc}
\usepackage[latin1]{inputenc}
\usepackage{color}
\usepackage{float}
\usepackage{amsmath}
\usepackage{amssymb}
\usepackage{graphicx}
\usepackage{esint}

\makeatletter
\@ifundefined{textcolor}{}
{%
 \definecolor{BLACK}{gray}{0}
 \definecolor{WHITE}{gray}{1}
 \definecolor{RED}{rgb}{1,0,0}
 \definecolor{GREEN}{rgb}{0,1,0}
 \definecolor{BLUE}{rgb}{0,0,1}
 \definecolor{CYAN}{cmyk}{1,0,0,0}
 \definecolor{MAGENTA}{cmyk}{0,1,0,0}
 \definecolor{YELLOW}{cmyk}{0,0,1,0}
}

\@ifundefined{definecolor}
 {\usepackage{color}}{}
\@ifundefined{definecolor}
 {\@ifundefined{definecolor}
 {\usepackage{color}}{}
}{}
\@ifundefined{definecolor}
 {\@ifundefined{definecolor}
 {\@ifundefined{definecolor}
 {\usepackage{color}}{}
}{}
}{}
\@ifundefined{definecolor}
 {\@ifundefined{definecolor}
 {\@ifundefined{definecolor}
 {\@ifundefined{definecolor}
 {\usepackage{color}}{}
}{}
}{}
}{}
\makeatother

\makeatother

\usepackage{babel}

\makeatother

\usepackage{babel}

\makeatother

\usepackage{babel}

\makeatother

\usepackage{babel}
\renewcommand\[{\begin{equation}}
\renewcommand\]{\end{equation}}
\makeatletter

\@ifundefined{showcaptionsetup}{}{%
 \PassOptionsToPackage{caption=false}{subfig}}
\usepackage{subfig}
\makeatother

\usepackage{babel}

\begin{document}

\title{Quantum criticality at the Anderson transition: a TMT perspective}

\author{Samiyeh Mahmoudian}

\affiliation{Department of Physics and National High Magnetic Field Laboratory,
Florida State University, Tallahassee, Florida 32306, USA.}

\author{Shao Tang}

\affiliation{Department of Physics and National High Magnetic Field Laboratory,
Florida State University, Tallahassee, Florida 32306, USA.}

\author{Vladimir Dobrosavljevi\'{c}}

\affiliation{Department of Physics and National High Magnetic Field Laboratory,
Florida State University, Tallahassee, Florida 32306, USA.}
\begin{abstract}
We present a complete analytical and numerical solution of the Typical
Medium Theory (TMT) for the Anderson metal-insulator transition. 
This approach self-consistently calculates the typical amplitude
of the electronic wave-functions, thus representing the conceptually simplest
order-parameter theory for the Anderson transition. We identify all possible
universality classes for the critical behavior, which can be found
within such a mean-field approach. This provides insights into how
interaction-induced renormalizations of the disorder potential may
produce qualitative modifications of the critical behavior. We also formulate
a simplified description of the leading critical behavior,
thus obtaining an effective Landau theory for Anderson localization. 
\end{abstract}
\maketitle

\section{Introduction}

Many physical systems display puzzling features, which are often associated with
the metal-insulator transition (MIT) \cite{dobrosavljevic2012conductor}.
Although the important roles of both the Anderson \cite{anderson1958absence}
(disorder-driven) and the Mott \cite{mott-book90} (interaction-driven)
routes to localization have been long appreciated, formulating a simple
order-parameter theory describing their interplay has remained a challenge.
Important advances have been achieved, over the last twenty years,
with the development of Dynamical Mean Field Theory (DMFT) \cite{georges1996dynamical}
methods, which provided new insights into how such an order-parameter
theory can be constructed. Although the original DMFT formulation
adequately describes many features of strongly correlated electron
systems, it proved unable to capture Anderson localization
effects, which cannot be neglected in presence of sufficiently strong
disorder \cite{RoP2005review}. 

To overcome these limitations, DMFT was extended to describe spatially
nonuniform systems, in approaches sometimes called ``Statistical
DMFT'' \cite{RoP2005review,dobrosavljevic1997mean,mirandavlad1,aguiaretal1,atkinson2007prb,tran2007prb,atkinson2009jpc,andrade09prl,hofstetter2010prb-a,hofstetter2010prb-b,hofstetter2011prb,aguiar2013prl}
(some authors call the same approach ``Real-Space DMFT'' \cite{potthoff1999prb,rosch2008prl,hofstetter2008njp}).
Here, the local DMFT order parameters (i.e. the appropriate local
self-energies) are self-consistently calculated at each lattice site
for a given realization of disorder, in a fashion similar to the Thouless-Anderson-Palmer
(TAP) theory \cite{TAP} for spin glasses. These efforts immediately
produced a wealth of new information, discovering phenomena such as
disorder-driven non-Fermi liquid behavior \cite{RoP2005review} and
the emergence of Electronic Griffiths Phases \cite{andrade09prl,dobrosavljevic2012conductor}
in the vicinity of the MIT. Despite these advances, progress has remained
slow, primarily because such approaches typically require very large-scale
numerical computations.

The missing key point in all these formulations was
the lack of an appropriate \emph{local} order parameter, which is capable
of recognizing Anderson localization. A hint on how to overcome this
difficulty was first provided in the seminal 1958 work by P. W. Anderson
\cite{anderson1958absence}, who emphasized that the \emph{typical}
(i.e. geometrically averaged) local density of states (TDOS) vanishes
at the transition, in contrast to its algebraically averaged counterpart.
This idea was later confirmed by large-scale computational studies
\cite{re:Janssen98} of the wave-function amplitude statistics, which
suggested that this quantity should play the role of an appropriate
order-parameter for this problem. 

A self-consistent calculation of TDOS was recently formulated, dubbed
``Typical-Medium Theory'' (TMT) \cite{pastor2001tmt}, which can
be regarded as the conceptually simplest order-parameter approach
for Anderson localization. This method uses the same ``cavity-field''
construction as in standard DMFT methods \cite{georges1996dynamical},
and represents an elegant and effective approach to treat both the
correlation and the localization effects on the same footing. Following
its discovery in 2003, TMT was quickly applied to various problems
with both interactions and disorder \cite{hofstetter2005prl,aguiar2006prb,hofstetter2009prl,aguiar2009prl,aguiar2013prl,aguiar2014prb},
providing useful new information which would be difficult to obtain
by alternative methods. The numerical solution of TMT equations has
been obtained for both the (non-interacting) Anderson \cite{pastor2001tmt,dobrosavljevic2010typical},
and the Mott-Anderson \cite{hofstetter2005prl,aguiar2006prb,aguiar2009prl,hofstetter2009prl,aguiar2014prb}
transition. However, deeper understanding of what one can generally
expect from TMT approaches would require a complete
analytical solution for the critical behavior, which has not been
available so far. 

Further motivation for our work is found in recent experiments that
were able to visualize the electronic wave function near the metal-insulator
transition, via scanning tunneling microscopy on $\mathrm{Ga_{1-x}Mn_{x}As}$
\cite{richardella2010visualizing}. This work highlighted the crucial
importance of the long-range Coulomb interaction, and confirmed the
early theoretical prediction of Efros and Shklovskii (ES) \cite{efros1975coulomb,efros1976coulomb},
that Coulomb interactions lead to the formation of a pseudogap within
the insulating phase. Within the ES picture, the gap opening is produced
by the electrostatic shifts of the (random) site energies, resulting
in a significantly renormalized probability distribution for the effective
random potential seen by the electrons. While the ES mechanism is
by now well documented by both theoretical and experimental studies
on the insulating side of the MIT \cite{efros1985electron}, its precise role
for the critical region has remained elusive. At the minimum, one should
investigate the effects of such pseudo-gap opening in the form of
the distribution function for disorder, and its role at the Anderson
transition. 

In this paper, we address and clearly answer the following physical
questions: (1) What types of quantum criticality can be found, for
the noninteracting Anderson localization transition, within the TMT
scheme, and how does the result depend on the model dependent details
of the band structure (e.g. particle-hole symmetry)? (2) How is the
critical behavior modified in cases where the renormalized
disorder distribution assumes a pseudo-gap form predicted by the ES
theory? We accomplish this by first presenting a detailed numerical
solution of the TMT equation, for several cases of relevance. We then
obtain a full analytical solution of the TMT equation, describing
the leading critical behavior which is in complete agreement with
the numerics, and includes the emergence of logarithmic corrections
to scaling. This insight is shown to provide a new perspective and
a simple physical understanding of several puzzling features of the
critical behavior, previously observed in both numerical studies and
in experiments. 

The rest of the paper is organized as follows. In section II we present
the general formulation of Typical-Medium Theory, and provide some
illustrative examples of relevance to experiments. We show that two
distinct types of critical behavior can be found within TMT, and investigate
their main features. A general strategy to analytically solve the
critical behavior within TMT is discussed in Section III, based on
an expansion in powers of order parameter (TDOS). We explain why a
simple solution can be obtained only in the special case of particle-hole
symmetry, which already provides a classification of possible types
of quantum criticality within TMT. We further investigate  
how it is affected by the form of distribution of random site energies.
In section IV we present a detailed analytical solution for the leading
critical behavior in absence of particle-hole symmetry, by reducing
the problem to a close-form solution of an appropriate Fredholm integral
equation. We show that particle-hole asymmetry leads to the emergence
of logarithmic corrections to scaling, leading to a (mild) modification
of the critical behavior at the mobility edge away from the band center.
Finally, based on our full understanding of the mathematical structure
of the theory, we present a simplified Landau theory
for Anderson localization in Section V. This approximation ignores the relatively mild logarithmic
corrections, but is still shown to capture all the important qualitative
trends of the full TMT solution, and to reproduce most of the qualitative
features observed in the large-scale numerics, as well as in some
experiments.

\section{Model and numerical solution of TMT equations}

The general strategy in formulating a local order-parameter theory
such as TMT follows the ``cavity'' method typically used in Dynamical
Mean Field Theory approaches \cite{georges1996dynamical}. Here, the
dynamics of an electron on a given site can be obtained by integrating
out all the other sites, and replacing its environment by an appropriately
averaged ``effective medium'' characterized by a local self energy
$\Sigma(\omega)$. This method can be utilized to self-consistently
calculate any desired local quantity, and in the following we briefly
review its application to TMT of Anderson localization \cite{pastor2001tmt,dobrosavljevic2010typical}.
For simplicity, we concentrate on a single band tight binding model
of non-interacting electrons with random site energies $\epsilon_{i}$
with a given distribution \textcolor{black}{$P(\epsilon_{i})$, which
the Hamiltonian of this system can be written as:}\textcolor{blue}{{} }

\[
H=\sum_{\left\langle ij\right\rangle ,\sigma}t_{ij}c_{i\sigma}^{\dagger}c_{j\sigma}+\sum_{i,\sigma}\varepsilon_{i}c_{i\sigma}^{\dagger}c_{i\sigma}.
\]
Here, $c_{i\sigma}^{\dagger}$ and $c_{i\sigma}$ are the electron
creation and annihilation operators, and $t_{ij}$ are the inter-site
hopping elements. The local (retarded) Green function corresponding
to site $i$ can be written as 
\begin{equation}
G_{ii}(\omega,\epsilon_{i})=[\omega+i\eta-\epsilon_{i}-\Delta(\omega)]^{-1},\label{eq:LocalGreen}
\end{equation}
where the ``cavity field'' $\Delta(\omega)$ represents the effective
medium, i.e. available electronic states to which an electron can
\textcolor{black}{hop from}\textcolor{red}{{} }of a given lattice site.
It is defined by incorporating the local self-energy $\Sigma(\omega)$
as 

\[
\Delta(\omega)=\Delta_{0}(\omega-\Sigma(\omega)),
\]
where $\Delta_{0}(\omega)$ is the ``bare'' (corresponding to zero
disorder) cavity field \cite{georges1996dynamical}. It can be obtained
from the bare lattice Green's function through\textcolor{red}{{} }\textcolor{black}{relation}
\begin{equation}
\Delta_{0}(\omega)=\omega-\frac{1}{G_{0}(\omega)},\label{eq:delta0(omega)}
\end{equation}
and\textcolor{red}{{} }\textcolor{black}{the bare lattice Green's function}
\begin{equation}
G_{0}(\omega)=\int_{-\infty}^{+\infty}d\omega^{\prime}\frac{\nu_{0}(\omega^{\prime})}{\omega+i\eta-\omega^{\prime}}.\label{eq:bare lattice Green's function}
\end{equation}
is given by the Hilbert transform of the bare density of states $\nu_{0}(\omega)$
\textcolor{black}{(DOS),} which specifies the electronic band structure
for a given lattice. The corresponding local density of states (LDOS)
is given by the imaginary part of the local Green's function: 
\begin{equation}
\rho_{i}(\omega,\epsilon_{i})=-\frac{1}{\pi}\mathrm{Im}G_{ii}(\omega,\epsilon_{i}).\label{eq:rho_i}
\end{equation}

Within the effective-medium approximation we consider, this local
quantity displays site-to-site fluctuations. Due to its dependence
on the local site energy $\epsilon_{i}$, it reflects the spatial
fluctuations of the local wave-function amplitudes $\rho_{i}\sim|\psi_{i}|^{2}$.
To properly define the effective medium, one has to perform an appropriate
spatial average, in order to close the self-consistency loop. The
simplest choice is to consider its algebraic average (ADOS)

\begin{equation}
\rho_{avg}(\omega)=\int d\epsilon_{i}P(\epsilon_{i})\rho_{i}(\omega,\epsilon_{i})\label{eq:rho_average}
\end{equation}
as the appropriate order parameter, and this leads to the well-known
coherent-potential approximation (CPA) \cite{Economou2006}, which
unfortunately fails to capture Anderson localization. 

In the presence of strong disorder, however, LDOS displays strong spatial
fluctuations and is very broadly distributed. As a result, its \emph{typical
(i.e., }most probable) value is ill-represented \cite{anderson1958absence}
by the algebraic average $\rho_{avg}(\omega)$. Since the average
density of states can remain finite throughout the insulating phase
(even in the atomic limit) as well as in the metallic phase,
it cannot distinguish between the phases. Therefore, within TMT,
we introduce the\emph{ typical value} of the local density-of-states,
as an appropriate order parameter. The statistic of LDOS reflects
the degree of localization of quantum wave functions, and its typical
value (TDOS) is known \cite{re:Janssen98} to be well-represented
by the geometric average

\begin{equation}
\rho_{typ}(\omega)=\exp\left[\int d\epsilon_{i}P(\epsilon_{i})\ln\rho_{i}(\omega,\epsilon_{i})\right].\label{eq:typical density}
\end{equation}
Indeed, large-scale computational studies, as well as the available
analytical results in $d=2+\varepsilon$ dimensions , demonstrated
that TDOS vanishes in a power-law fashion at the critical point, and
also displays the appropriate finite-size scaling behavior (for reviews
see Refs. \cite{re:Janssen98,evers2008anderson}). These results strongly
suggest \cite{re:Janssen98} that TDOS should be chosen as an appropriate
local order parameter; its self-consistent calculation can be viewed
as the conceptually simplest order-parameter theory of Anderson localization.
In order to obey causality, the corresponding \emph{``typical''
}Green's function, is defined \cite{pastor2001tmt,dobrosavljevic2010typical}
by performing the Hilbert transform
\begin{equation}
G_{typ}(\omega)=\int_{-\infty}^{\infty}d\omega^{\prime}\frac{\rho_{typ}(\omega^{\prime})}{\omega+i\eta-\omega^{\prime}}.\label{eq:Gtyp}
\end{equation}
Note that the $G_{typ}(\omega)$ has to be defined
on the real frequency axis, because this is computed where LDOS is
defined as a positive definite quantity and has a well-defined geometric
average. Finally, we close the self-consistency loop \textcolor{black}{by
setting the Green functions of the effective medium to be equal to
that corresponding to the local order parameter} \cite{pastor2001tmt,dobrosavljevic2010typical},
\begin{equation}
G_{typ}(\omega)\equiv G_{0}(\omega-\Sigma(\omega)).\label{eq:selfconditionforGtyp}
\end{equation}
From this self-consistency condition and Eq.
(\ref{eq:delta0(omega)}), we obtain the following equation which
determines the self-energy of the system
\begin{equation}
G_{typ}(\omega)=[\omega+i\eta-\Sigma(\omega+i\eta)-\Delta(\omega+i\eta)]^{-1}.\label{eq:self energy}
\end{equation}

It is important to emphasize that our procedure defined by TMT self-consistent
equations (\ref{eq:LocalGreen}-\ref{eq:self energy}) is not specific
to the problem at hand; the same strategy is used in any mean-field
(DMFT-like) theory characterized by a local self-energy \cite{georges1996dynamical}.
The only requirement specific to TMT is the choice of the \emph{typical
}(geometrically-averaged) LDOS as the local order parameter. In other
words, the only crucial difference between CPA and TMT is the fact
that TMT utilizes the appropriate order parameter for Anderson localization.

This set of TMT self-consistent equations can be solved numerically
for any specific lattice model, or any form of the random site energy
distribution. However, as in any other mean-field formulation, only
a limited number of \emph{qualitatively} distinct types of critical
behavior (i.e., universality classes) can arise, and in the following
we discuss two distinct situations that we have found within TMT.
Previous work mostly focused on models with continuous (e.g., uniform)
distributions of site energies, and even some analytical results were
obtain in this case \cite{pastor2001tmt}.

In the following, we present the
results obtained numerically by solving the TMT equations for semi-circular
DOS which is given by  $\nu_{0}(\omega)=2\sqrt{1-\omega^{2}}/\pi$. Here and in 
the the rest of the paper, all energies are expressed
in units of the half-bandwidth.
As an example, we consider the uniform model where
the distribution of random site energies is continuous and is given
by $P_{uniform}(\epsilon_{i})\equiv\frac{1}{W},$
over the interval $-\frac{W}{2}\leqslant\epsilon_{i}\leqslant\frac{W}{2}$. We display the resulting 
behavior for this model \cite{pastor2001tmt} in Fig.~1, showing
the evolution of $\rho_{typ}(\omega)$ as
disorder increases. The extended states are identified by  the frequency
range where $\rho_{typ}>0,$ which is seen to shrink and eventually
disappear at a critical disorder $W=W_{c},$ where the entire band
localizes. The metallic phase is
separated from the Anderson insulator\textcolor{red}{{} }\textcolor{black}{insulating
phase} by the mobility edge trajectory $\omega=\omega_{c}(W)$, corresponding
to TDOS vanishing.

\begin{figure}[H]
\subfloat[]{\includegraphics[width=3in]{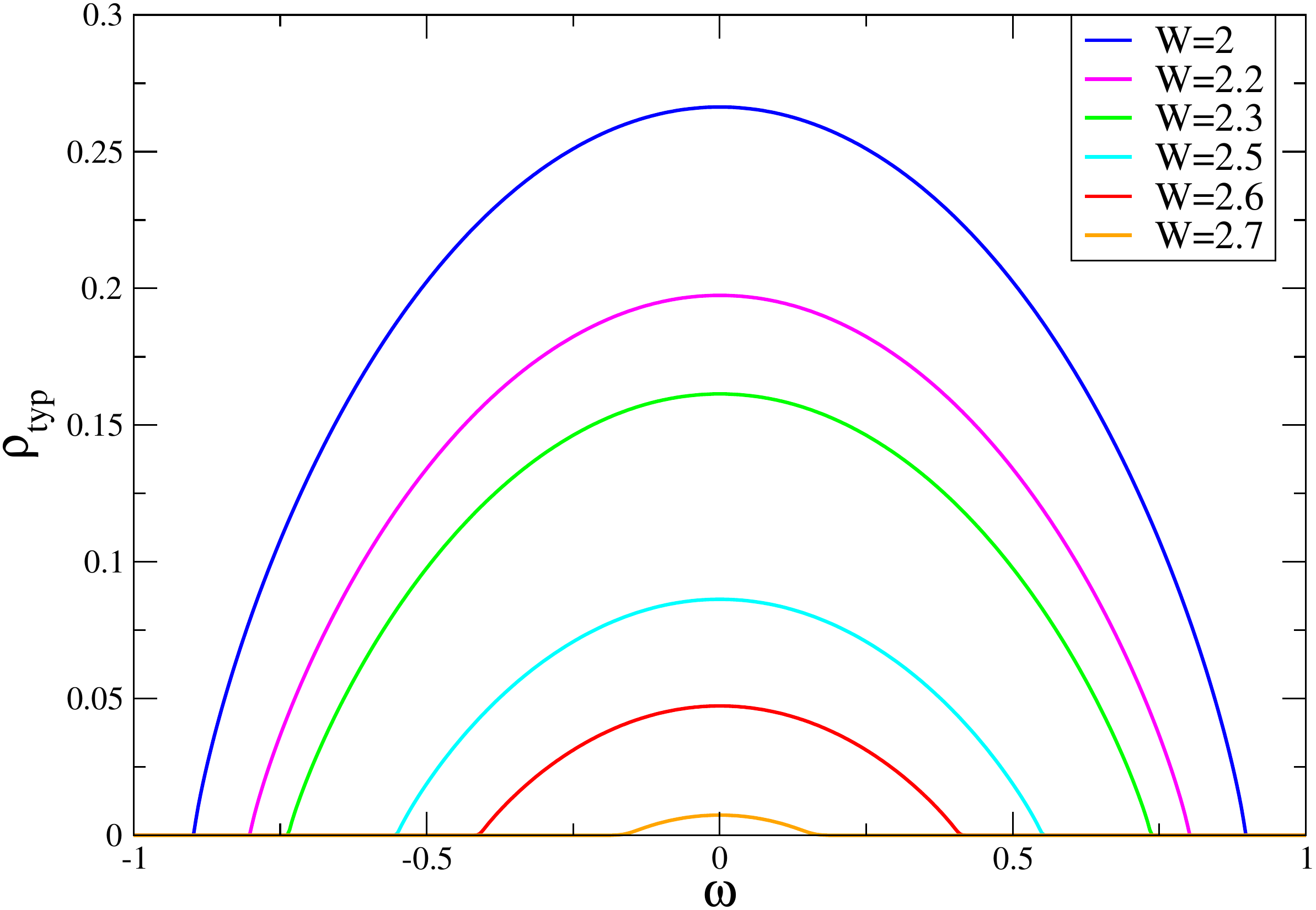}}

\subfloat[]{\includegraphics[width=3in]{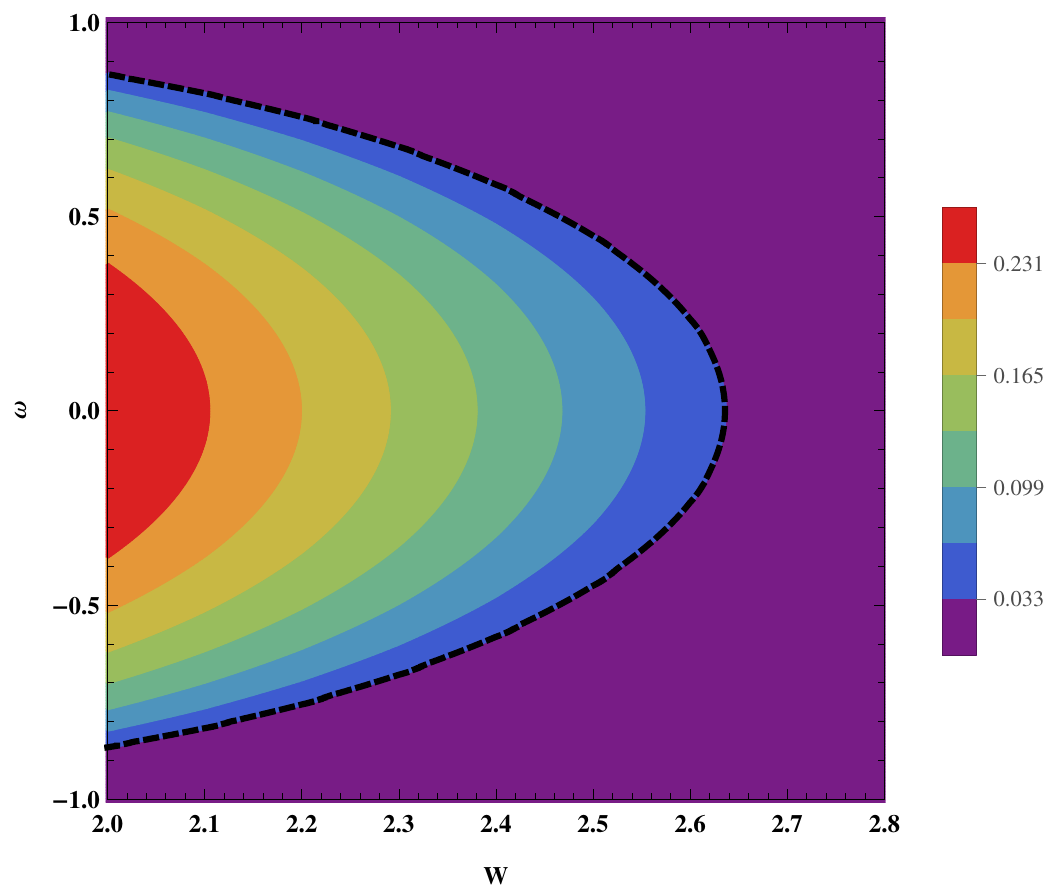}

}\protect\caption{Evolution of the order parameter $\rho_{typ}(\omega)$ with increasing
disorder for uniform model: $\rho_{typ}(\omega)$for several values of the disorder strength (top panel). The
bottom panel shows the phase diagram in the $\omega-W$ plane, where the
mobility edge (black dashed line) separates the extended states ($\rho_{typ}>0$)
from the localized states ($\rho_{typ}=0$), and $\rho_{typ}$ is color
coded. }
\end{figure}

The situation is qualitatively different if the disorder distribution
has a gap or a pseudo-gap, so that $P(\epsilon)$
vanishes at one energy or in an entire energy interval. This situation
can arise for discrete (e.g., binary) distributions of disorder, which
can be found in alloys. A similar situation can also arise in presence
of electron-electron interactions which we 
 briefly discuss in the following. Here, the \emph{effective} disorder
potential (i.e., the renormalized random potential) seen by quasi-particles
can be significantly modified by interaction effects, especially in
presence of \emph{long-range }Coulomb interactions, which
leads to the formation of the soft \emph{``Coulomb gap''
} (pseudo-gap) at the Fermi energy. This behavior, which was recently brought to
attention by scanning tunneling microscopy (STM) experiments \cite{richardella2010visualizing}
on $\mathrm{Ga_{1-x}Mn_{x}As}$, has been first discussed in the well-known
theoretical work of Efros and Shklovskii (ES) \cite{efros1975coulomb,efros1976coulomb,efros1985electron}.
These authors argued that the key effect of the long-range Coulomb
interactions is to provide a strong renormalizations of the electronic
on-site energies, due to the fluctuating electrostatic potential produced
by distant charges. \textcolor{black}{Therefore, the renormalized
site energy $\tilde{\varepsilon}_{i}$ is given by}

\[
\tilde{\varepsilon}_{i}=\varepsilon_{i}+e^{2}\sum_{j}\frac{n_{j}}{R_{ij}},
\]
where $\tilde{\varepsilon}_{i}$ is the renormalized site energy,
$n_{j}=0,1$ is the occupation number of a given lattice site $j$, $e$ is
the electron charge, and $R_{ij}$ is the distance between sites $i$
and $j$. 

According to the ES theory, the main result of the Coulomb interactions
is to produce a renormalized distribution of disorder, which (in spatial
dimension $d$) assumes a low-energy pseudo-gap form (vanishes in
power-law fashion) 
\[
P(\tilde{\varepsilon})\sim\tilde{\varepsilon}^{d-2},
\]
where the renormalized energy $\tilde{\varepsilon}_{i}$ is measured
with respect to the Fermi energy. In other words, the renormalized
distribution function vanishes at the Fermi energy, i.e., $P(0)=0,$
a situation which, as we shall see, leads to qualitatively different
critical behavior of TDOS within TMT. The ES result was derived using
a classical electrostatic model, which should be sufficient deep in
the Anderson-localized phase. Closer to the MIT, the precise form
of $P(\tilde{\varepsilon})$ may be affected by quantum fluctuations,
as argued in Ref. \cite{massey1996prl}, and it may need to be self-consistently
calculated, in order to accurately capture the
interplay of Anderson localization and the  effects of the Coulomb interactions.
Such a calculation may be possible within the framework of a DMFT-like
formulation, by combining TMT with the EDMFT approach to Coulomb correlations
\cite{2011prb}, but this rather complicated analysis is left as a challenge for future
work.

In this paper, we limit our attention to analyzing, within TMT,
the consequences of having such a pseudo-gap form for the disorder
distribution function. As an illustration, we consider a model distribution
of random site energies which assumes a pseudo-gap form expected from
the ES picture in three dimensions:

\begin{equation}
P_{pseudo}(\tilde{\varepsilon}_{i})\equiv\frac{1}{(\frac{W}{6})^{3}\sqrt{2\pi}}\tilde{\varepsilon}_{i}^{2}\exp(-\frac{\tilde{\varepsilon}_{i}^{2}}{2(\frac{W}{6})^{2}}),\label{eq:P_pseudogap}
\end{equation}
which we will refer as pseudo-gap model\footnote{The value of W  for the pseudogap model is normalized in such a way
that $<\varepsilon_{i}^{2}>$ has the same value for both the uniform and the pseudogap
models.} in the following text. We
solved the TMT equations for this model of disorder, and the results
for $\rho_{typ}(\omega)$ and $\rho_{avg}(\omega)$ are presented in
Figs. 2 and 3. As disorder increases, the TDOS order parameter
displays the most pronounced decrease precisely at the Fermi energy (here
chosen at $\omega=0$); the corresponding electronic state is the
one to first localize at the critical disorder strength $W_{c1}=2.07$.
As disorder increases further, there emerges a finite ``mobility
gap'' around the Fermi energy, where our TDOS order parameter $\rho_{typ}$
vanishes at $|\omega|<\omega_{c}(W)$, and all the electronic states
\textcolor{black}{within this region} become localized. At even larger
disorder $W=W_{c2}$ the entire band localizes. The trajectories of
the corresponding mobility edges (shown by a dashed black line in
Fig. 2(b)) displays the same non-monotonic behavior as found in the recent
large-scale numerical study of the localization transition in 
Coulomb glasses \cite{amini2013multifractality}.

\begin{figure}[H]
\subfloat[]{\includegraphics[width=3in]{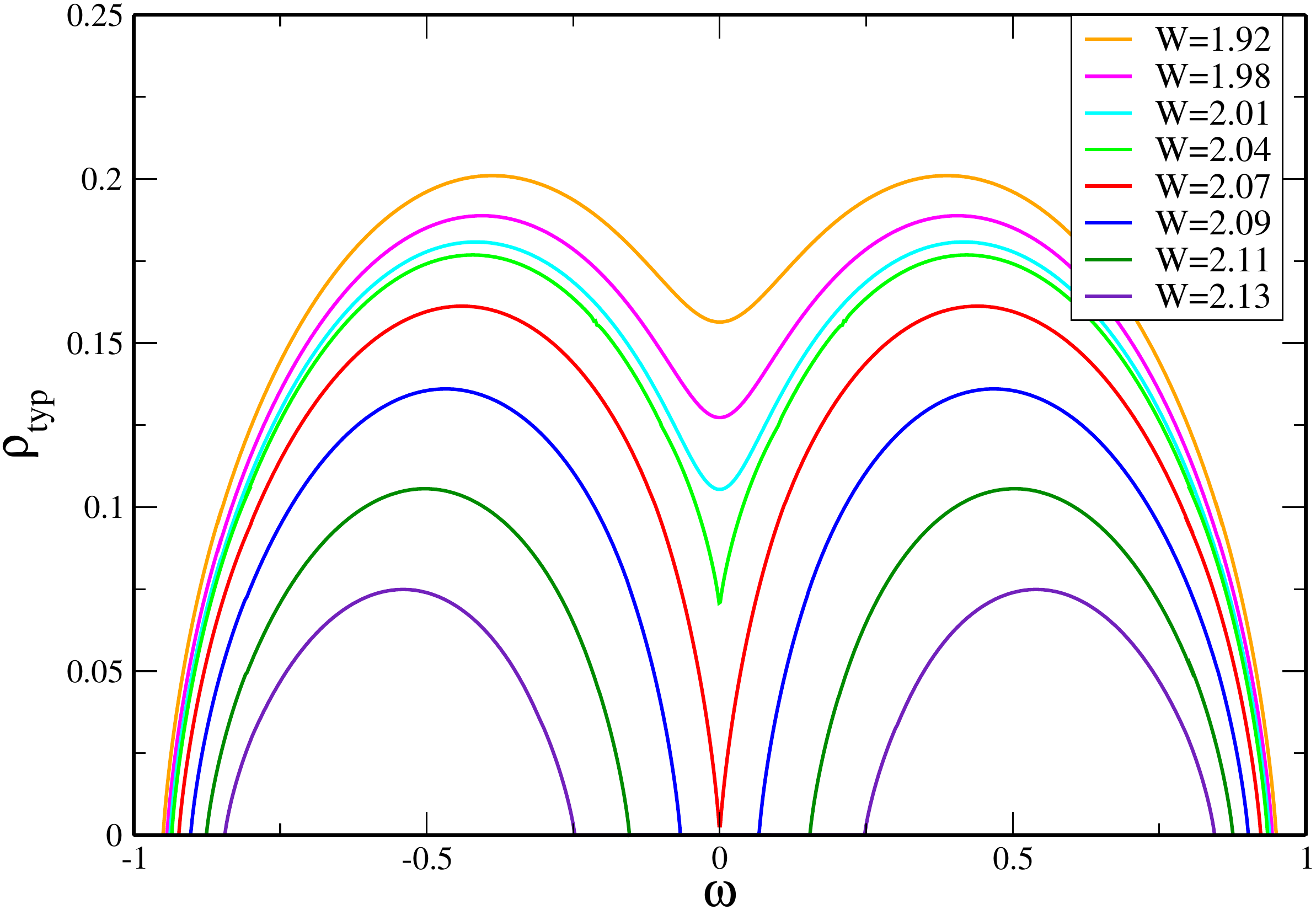}}

\subfloat[]{

\includegraphics[width=3in]{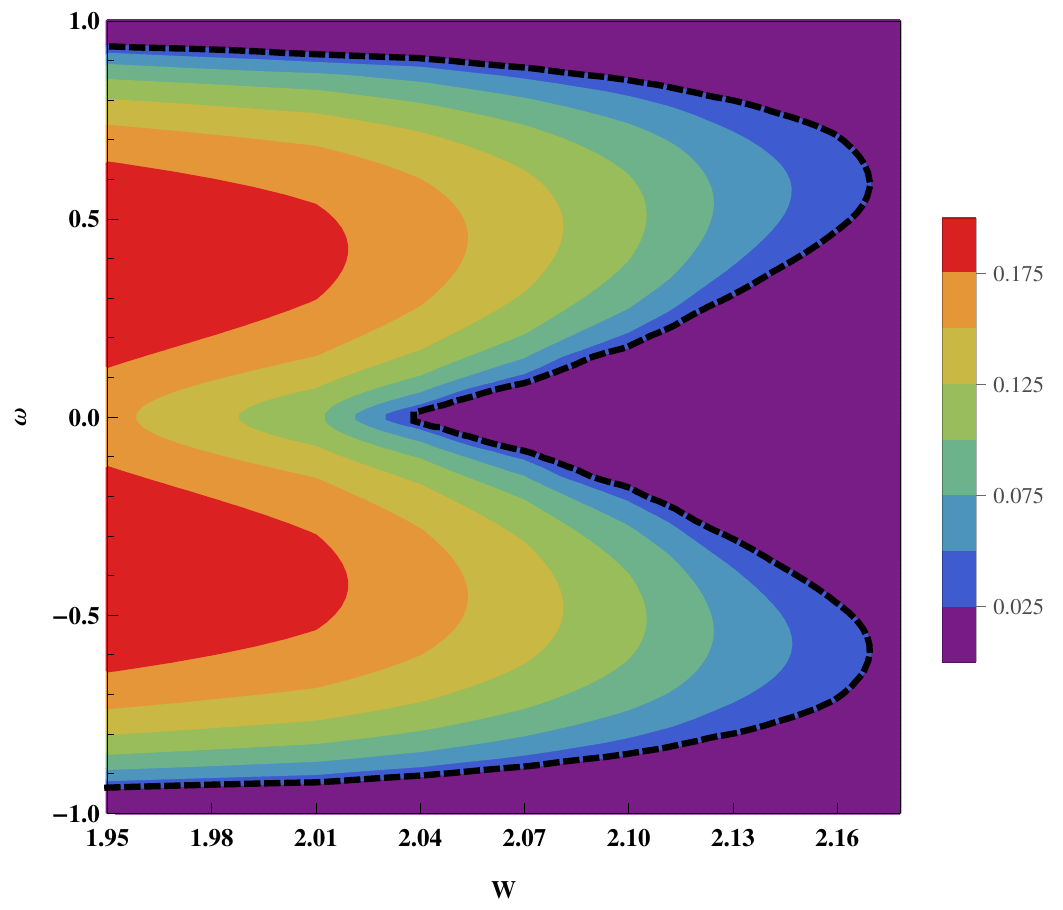}

}

\protect\caption{Evolution of the order parameter $\rho_{typ}(\omega)$ with increasing
disorder for pseudo-gap model: $\rho_{typ}(\omega)$\textcolor{red}{{}
}\textcolor{black}{for several disorder strength} (top panel).\textcolor{black}{{}
The electronic states near the Fermi energy ($\omega=0$) are localized
before the rest of the band would be localized. }}
\end{figure}

For comparison with experiments, we also computed the algebraically-averaged
local density of states (ADOS), which shows very different behavior.
ADOS at the Fermi energy ($\omega=0$ ) is found to vanish at precisely
the same critical disorder $W=W_{c1}$ for localization \cite{richardella2010visualizing},
but it remains finite at all other energies ($|\omega|>0$) within
the localized phase, as shown in Fig. 3(b). Since we found  that 
TDOS vanishes for $W>W_{c1}$ and $|\omega|<\omega_{c}(W)$
in Fig. 3(a), our numerical results immediately reveal that, within
the entire localized phase, ADOS assumes a power-law low energy form 
\begin{equation}
\rho_{avg}(\omega)\sim\omega^{2}.\label{eq:Ef_behavior}
\end{equation}
In order to analytically understand this result, note that from Eq. (7), ADOS can be expressed as:
\[
\rho_{avg}(\omega)=\frac{1}{\pi}\int d\epsilon P(\epsilon)\frac{\Delta^{\prime\prime}(\omega)}{(\omega-\varepsilon-\Delta^{\prime}(\omega))^{2}+\Delta^{\prime\prime}(\omega)^{2}}.
\]
At $W>W_{c1}$ the imaginary part of the cavity field also vanishes
at region $|\omega|<\omega_{c}(W),$ since it behaves as $\Delta^{\prime\prime}\sim\rho_{typ}$
(See appendix A). As it can be proven straightforwardly
and is also shown numerically, the real part of the cavity field
is a linear function as $\Delta^{\prime}(\omega)=A\omega$
with $A$ a finite constant, and we find
\begin{eqnarray*}
\rho_{avg}(\omega) & = & \frac{1}{\pi}\lim_{\Delta^{\prime\prime}\rightarrow0}\{\int d\epsilon P(\epsilon)\frac{\Delta^{\prime\prime}}{((1-A)\omega-\varepsilon)^{2}+\Delta^{\prime\prime2}}\}\\
 & = & P((1-A)\omega)\sim P(\omega)\sim \omega^2,
\end{eqnarray*}
in agreement with ES theory.

Our  results thus provide a qualitative picture
of pseudo-gap formation of $\rho_{avg}(\omega)$, which is centered
at $\omega=0$ both at the critical point $(W=W_{c1})$ and in the entire insulating
phase ($W>W_{c1}$). This result (also shown in Fig. 3(b)) is consistent
with large-scale exact diagonalization results \cite{amini2013multifractality}, and the available
experimental findings \cite{massey1996prl,richardella2010visualizing}.

The emergence of qualitative different critical behavior, for the two distinct
models of disorder, is even more clearly seen by examining our
order parameter $\rho_{typ}$ at the center of the band $(\omega=0)$.
Fig. 4 (top panel) shows that for pseudo-gap model  $\rho_{typ}$  vanishes
as square root of distance from transition viz.  $\rho_{typ}\sim(W_{c1}-W)^{\frac{1}{2}}$,
while for the uniform model (Fig. 4  bottom panel) we find linear
behavior viz. $\rho_{typ}\sim(W_{c}-W)$.
In order to try and understand the origin of these differences, 
in Section III we analytically recover the same critical behaviors  at the 
band center $(\omega=0)$.
Although this result gives us insight into the important differences between the two models, away from the band center this behavior cannot be explained in a simple way. This fact has been identified in previous work\cite{dobrosavljevic2010typical}; it has so far remained ill-understood, and clarifying this issue 
is the subject of our complete analytical solution in Sec. IV.
\begin{figure}[H]
\subfloat[]{\includegraphics[width=3.5in]{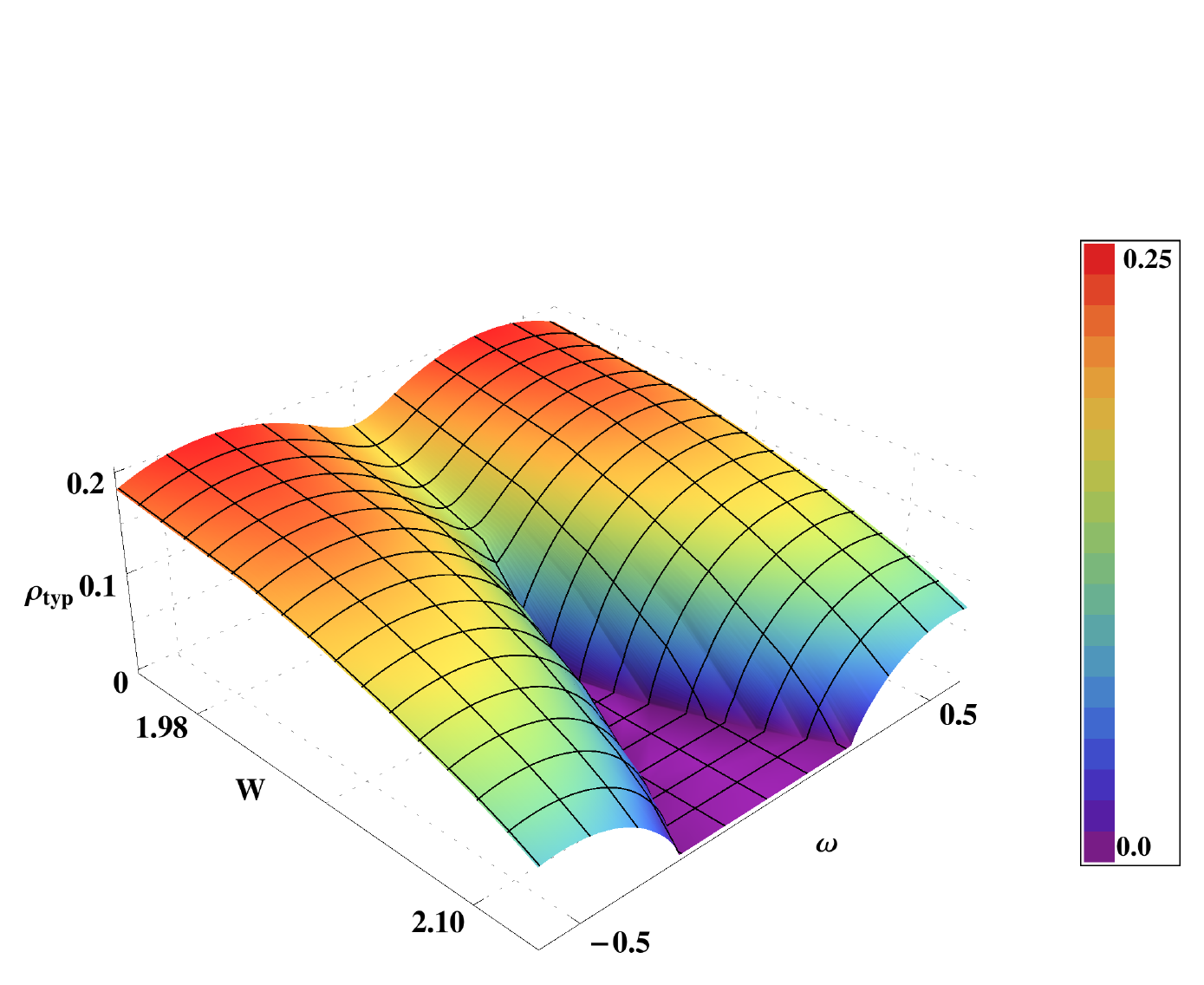}}

\subfloat[]{

\includegraphics[width=3.5in]{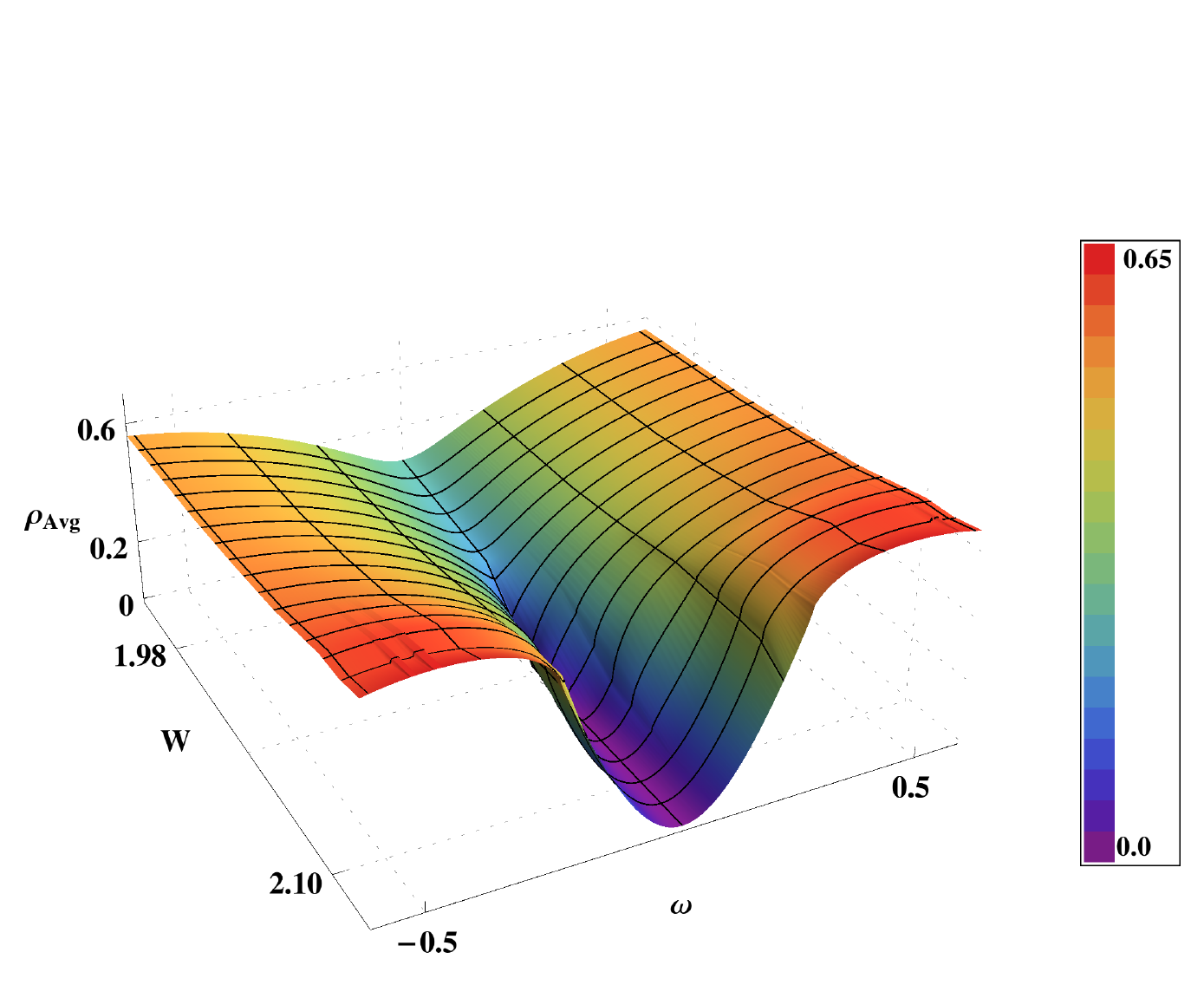}

}\protect\caption{The evolution of the (a) typical and (b) average density of states
for pseudo-gap model.}
\end{figure}

\section{Analytical solution: the Landau expansion}

It is well known that the Anderson transition is a second-order phase
transition, where the order parameter $\rho_{typ}(\omega)$ vanishes
continuously as the transition is approached, as also confirmed by our numerical solution of the TMT equations. 
Using the fact that $\rho_{typ}(\omega)$ is infinitesimally small in the close
vicinity of the transition, we can proceed
as in deriving any Landau theory, by directly expand the TMT
equations in the powers of the order parameter. For the sake of simplicity
in notation we define 
\[
\rho_{typ}(\omega)\equiv\varphi(\omega).
\]
 The Anderson transition is found along the critical (mobility edge) line on the phase diagram, defined by the expression
\[
\varphi[\omega_{c}(W)]=0,
\]
as shown by a black dashed line in Fig. 1(b) and Fig. 2(b). 

In order to obtain the solution as
the transition is approached, we start with the general expression for TDOS, as given by Eq.(\ref{eq:typical density}),
Eq.(\ref{eq:rho_i}), and Eq.(\ref{eq:LocalGreen}), which can be rewritten as
\begin{equation}
\pi\varphi(\omega)=\Delta^{\prime\prime}(\omega)g(\Delta^{\prime\prime}(\omega),\Delta^{\prime}(\omega)),\label{eq:g-omega}
\end{equation}
where,
\begin{eqnarray*}
g(\Delta^{\prime\prime}(\omega),\Delta^{\prime}(\omega)) & \equiv & \exp\{-\int d\epsilon P(\epsilon)\\
 &  & \times\log[(\omega-\epsilon-\Delta^{\prime}(\omega))^{2}+\Delta^{\prime\prime}(\omega)^{2}]\}.
\end{eqnarray*}

To proceed, we note that near the mobility edge, where $\varphi\ll1,$
the imaginary part of the cavity field is also small $(\Delta^{\prime\prime}\ll1)$,  
since to leading order \cite{pastor2001tmt} 
\begin{equation}
\Delta^{\prime\prime}=C\pi\varphi,\label{eq:general_cavity_halffill}
\end{equation}
where $C=\int d\omega^{\prime}\omega^{\prime2}\nu_{0}(\omega^{\prime})$,
and $\nu_{0}(\omega)$ is a bare density of states (See Appendix A). In contrast, 
 $\Delta^{\prime}(\omega_{c})$ generally remains finite. Indeed, we checked numerically
 that all qualitative features of the critical behavior do not depend on the specific choice of band structure \cite{pastor2001tmt},  which only modifies the precise 
value of the prefactor $C$ in Eq.(\ref{eq:general_cavity_halffill}), and other non-universal quantities. 
We can, therefore, expand the right hand side of Eq.(\ref{eq:g-omega})
in terms of $\varphi\sim \Delta^{\prime\prime}$, giving us a Landau-type expansion of the form
\begin{equation}
\frac{1}{C}=[a(\omega)+b(\omega)\varphi+d(\omega)\varphi{}^{2}+...].\label{eq:expansion 3}
\end{equation}
Here,
\begin{equation}
a(\omega)\equiv\exp\{-2\int P(\epsilon)d\epsilon\log\mid\omega-\epsilon-\Delta^{\prime}(\omega)\mid\},\label{eq:a(omega)}
\end{equation}
\begin{equation}
b(\omega)=-2\pi^{2}a(\omega)P(\omega-\Delta^{\prime}(\omega)),\label{eq:B(omega)}
\end{equation}
\[
d(\omega)=a(\omega)(\eta\pi^{2}+2\pi^{4}P(\omega-\Delta^{\prime}(\omega))^{2}),
\]
and
\begin{equation}
\eta=\lim_{\Delta^{\prime\prime}\rightarrow0}\{\int d\epsilon P(\epsilon)\frac{(\Delta^{\prime\prime}{}^{2}-\left(\omega-\epsilon-\Delta^{\prime}(\omega)\right)^{2})}{(\left(\omega-\epsilon-\Delta^{\prime}(\omega)\right)^{2}+\Delta^{\prime\prime}{}^{2})^{2}}\},\label{eq:eta}
\end{equation}
where $\eta$ remains finite for the models we examined. 
\begin{figure}[H]
\subfloat[]{\includegraphics[angle=270,width=3.5in]{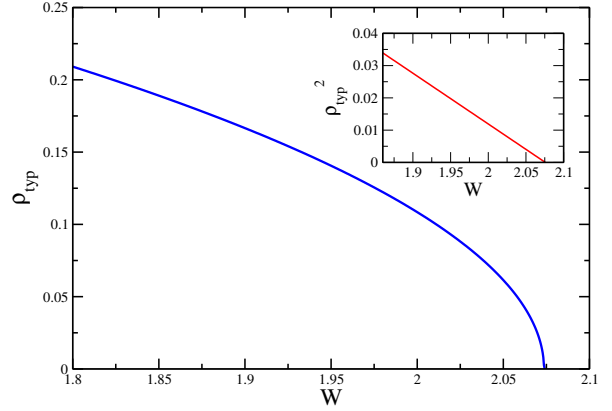}}

\subfloat[]{\includegraphics[width=3in]{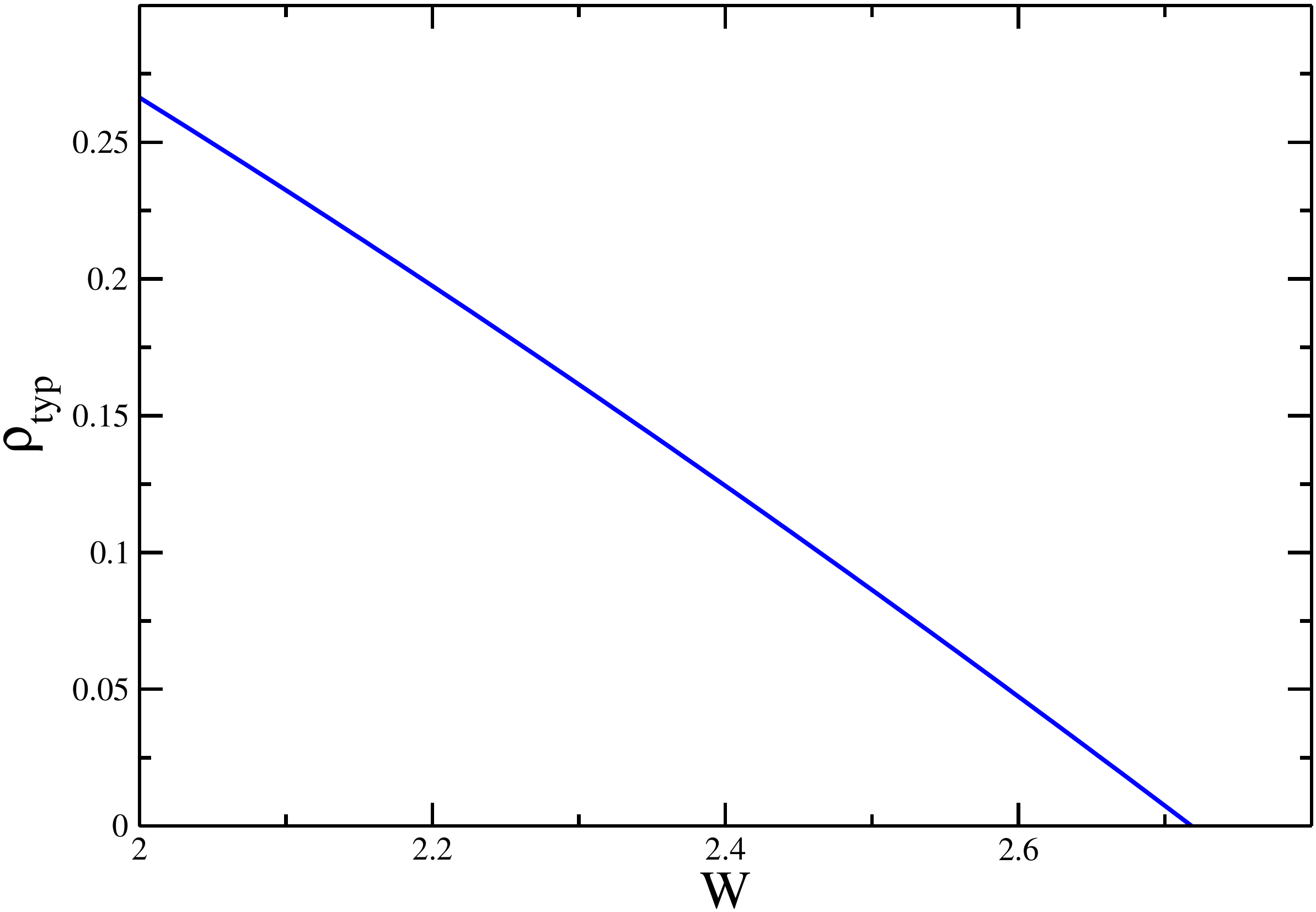}}

{\protect\caption{Critical behavior of TDOS as a function
of disorder strength $W$  at half-filling ($\omega=0$), for (a) the pseudo-gap model, and (b) the uniform
model of random site energies.}}
\end{figure}

\subsection{General critical behavior}

As in any Landau theory, we can now directly obtain the critical behavior
of the order parameter $\varphi(\omega )$, in terms of the coefficients in the expansion. 
For simplicity, consider a simple model band structure with
semi-circular DOS where $C=1$, and solve the Eq.
(\ref{eq:expansion 3}) for the order parameter $\varphi(\omega)$. 
For the generic model (e.g. uniform distribution of disorder) where $b(\omega_{c})\neq0$, the leading critical behavior of typical density of states takes the form
\begin{equation}
\varphi(\omega)=\frac{1-\frac{1}{a(\omega)}}{2\pi^{2}P(\omega-\Delta^{\prime}(\omega))},\label{eq:phi(omega_uniform)}
\end{equation}

In contrast, whenever
$b(\omega_{c})=0$ (e.g. the pseudo-gap model), we find
\begin{equation}
\varphi(\omega)=(\frac{\frac{1}{a(\omega)}-1}{\eta\pi^{2}})^{\frac{1}{2}},\label{eq:phi(omega_pseudogap)}
\end{equation}

At first glance, it seems that the critical behavior can be obtained
easily given the cavity field, which is the functional of order parameter
$\varphi(\omega),$ as $\Delta(\omega)=\mathcal{F}[\varphi(\omega)].$
However, the analytical solution is very complicated because the real
part of the cavity field $\Delta^{\prime}(\omega)$ is an unknown
function of\textcolor{red}{{} }\textcolor{black}{$\omega$} which is
linked to imaginary part $\Delta^{\prime\prime}(\omega)$ by the Hilbert
transform. Thus, it is impossible to solve these equations analytically
over a broad frequency range. As it has been shown numerically, we claim
that this unknown function is finite near the mobility edge $(\omega\approx\omega_{c})$.
 In order to obtain the leading critical behavior of order parameter at transition,  
 in in expression containing  $P(\omega)$, we can replace
\begin{equation}
P(\omega-\Delta^{\prime}(\omega))\approx P(\omega_{c}-\Delta^{\prime}(\omega_{c})).\label{eq:approximate P(omega)}
\end{equation}
However, in other terms, e.g. in the expression for $a(\omega)$ (Eq. 22)),
one needs to retain the full frequency dependence, which proves
to assume a sufficiently singular form to contribute to leading order 
(see below). As a result, the critical behavior becomes a more 
complicated form, as we shall see from the full analytical solution of Eq.(\ref{eq:expansion 3})
close to the mobility edge. The specific form of the analytical solution  has
been provided in section IV for the two different classes of random distributions
(uniform and pseudogap-gaussian), and it successfully has been compared with numerical
results.

\subsection{Critical behavior at half-filling}

Here, we explore the exact functional form of order parameter close
the transition, focusing on half-filling, where $\Delta^{\prime}(0)=0$.
In this case, there is no need to perform the Hilbert transform, so
all Landau coefficients can be evaluated in closed form a 
\begin{equation}
a(0)=\exp\{-2\int P(\epsilon)d\epsilon\log\mid\epsilon\mid\}\equiv a,\label{eq:a(0)}
\end{equation}
and 
\[
b(0)=-2\pi^{2}P(0).
\]
Our Landau-like expansion now takes  simple form
\begin{equation}
1=a[1-2\pi^{2}P(0)\varphi+(\eta\pi^{2}+2\pi^{4}P(0){}^{2})\varphi^{2}+...].\label{eq:order_parameter_Half_fil}
\end{equation}

Note that here the value 
of $P(0)$ plays an important role, and this what causes two different forms of criticality, 
for the generic model where $P(0)\neq0$, and for the 
pseudo-gap model where $P(0)=0.$ For the generic case, the TDOS vanishes linearly at the transition
\[
\varphi\sim(W-W_{c}),
\]
while for models with $P(0)=0$, the critical behavior assumes a square-root form 
\[
\varphi\sim(W-W_{c1})^{\frac{1}{2}}.
\]
We emphasize that the parameter $\eta$ remains finite at the transition, and can be directly calculated
for any specific form for $P(\varepsilon)$ from Eq. (25).
The condition $a=1$ directly gives us the critical value of disorder; for example,
for the considered pseudo gap model we obtain $W_{c1}=2.07$, in excellent agreement with numerical results shown
in Fig .4 (a) and Fig. 4(b). 

\section{FREDHOLM INTEGRAL EQUATION AND GENERAL SOLUTION}

Here, we obtain the full analytical solution for the critical behavior, valid even away from particle hole
symmetry, and for an arbitrary model of disorder. 

\subsection{Analytical solution for a "generic model" with  $b(\omega_{c})\protect\neq0$}

The critical behavior of our TDOS order parameter, is 
given by Eq.(\ref{eq:phi(omega_uniform)}), where it is expressed in terms of (the yet unknown)
function $\Delta'(\omega)$. Note, however, that (viz. Eq. 20) in the critical region 
$\varphi(\omega) \sim \Delta''(\omega)$ is linked through the Hilbert transform to  $\Delta'(\omega)$, since 
\[
\Delta^{\prime}(\omega)=H[\Delta^{\prime\prime}(\omega)].
\]
Both quantities, therefore, need to be self-consistently calculated, as we do in the following. To do this, we
express the all expressions in terms of $\Delta''(\omega)$; for simplicity we focus on the semi-circular band structure model where $C=1$, and we can write
\begin{equation}
\Delta^{\prime\prime}(\omega)=(1-\frac{1}{a(\omega)})\frac{1}{2\pi P(\omega-\Delta^{\prime}(\omega))}.\label{eq:cavity_finite bandedge}
\end{equation}
Using Eq. (28), to  leading order we find
\begin{equation}
\Delta^{\prime\prime}(\omega)\approx \delta a(\omega)\frac{1}{2\pi P(\omega_{c}-\Delta^{\prime}(\omega_{c}))},\label{eq: cavity approximate near MIT}
\end{equation}
where $\delta a(\omega)\equiv (1-\frac{1}{a(\omega)})$ can be directly computed as a variation of $a(\omega)$
from Eq. (\ref{eq:a(omega)}) giving
\begin{equation}
\delta a(\omega)\approx 2\int d\epsilon P(\epsilon)\frac{1}{\omega-\epsilon-\Delta^{\prime}(\omega_c)}(\delta\omega-\delta\Delta^{\prime}(\omega)).\label{eq: variation of a}
\end{equation}
Here, $\delta\omega\equiv\left|\omega-\omega_{c}\right|$ and  $\delta\Delta^{\prime}(\omega)\equiv\Delta^{\prime}(\omega)-\Delta^{\prime}(\omega_{c})$. Using  Eq.(\ref{eq: cavity approximate near MIT})
we get the following integral equation linking $\Delta'(\omega)$ and $\Delta''(\omega)$
\begin{equation}
\Delta^{\prime\prime}(\omega)=\varLambda_{0}(\delta\omega-\delta\Delta^{\prime}(\omega)).\label{eq:universal}
\end{equation}
Here, $\varLambda_{0}$ is a finite number, given by
\[
\varLambda_{0}=\frac{1}{\pi P(\omega_{c}-\Delta^{\prime}(\omega_{c}))}\int d\epsilon\frac{P(\epsilon)}{\omega_{c}-\epsilon-\Delta^{\prime}(\omega_{c})}.
\]
This result is valid for any (generic) model of disorder with $P(\omega_{c} -\Delta^{\prime}(\omega_{c}))\neq 0$.  
 More explicitly, Eq.(\ref{eq:universal}) can be rewritten as 
\begin{equation}
\Delta^{\prime\prime}(\omega)-\frac{\varLambda_{0}}{\pi}\int_{-\infty}^{\infty}d\omega^{\prime}\frac{\Delta^{\prime\prime}(\omega^{\prime})}{\omega^{\prime}-\omega}=\varLambda_{0}(\delta\omega-\Delta^{\prime}(\omega_{c})).\label{eq:cavityfredhom}
\end{equation}
This integral equation (\ref{eq:cavityfredhom}) can be recognized as the Fredholm Integral
Equation (FIE), which assumes the form
\begin{equation}
y(x)-\lambda\int_{-\infty}^{\infty}dt\frac{y(t)}{t-x}=f(x).\label{eq:fredhom}
\end{equation}
By comparison of Eq.(\ref{eq:cavityfredhom}) and Eq.(\ref{eq:fredhom})
it can be seen that in our case $y(x)=\Delta^{\prime\prime}(\omega)$, $\lambda=\frac{\varLambda_{0}}{\pi},$
and $f(x)=\varLambda_{0}(\delta\omega-\Delta^{\prime}(\omega_{c}))$. For completeness, we outline in the following the standard reasoning used in solving the FIE. 
It uses the fact that the Hilbert transform is a linear operator, with the additional property of being "idempotent", i.e. obeying  \textcolor{black}{and}
$H^{2}=-1$; this immediately gives us a hint how to solve it in closed form. We  first apply the  Hilbert transform on Eq.(\ref{eq:fredhom}), 
and we can write
\begin{equation}
\frac{1}{\pi}\int_{-\infty}^{\infty}dt\frac{y(t)}{t-x}+\lambda\pi y(x)=H[f(x)].\label{eq:Hilbertfredhom}
\end{equation}

Next, we use  Eq.(\ref{eq:fredhom}) and Eq.(\ref{eq:Hilbertfredhom})
to eliminate\cite{polyanin2008handbook}  the term with the integral and express $y(x)$ entirely in terms of $f(x)$ giving

\[
y(x)=\frac{1}{1+\pi^{2}\lambda^{2}}\left\{ f(x)+\lambda\int_{-\infty}^{\infty}\frac{f(t)dt}{t-x}\right\} .
\]
Applying this solution to our Eq.(\ref{eq:cavityfredhom}) we find
\begin{equation}
\Delta^{\prime\prime}(\omega)=\frac{\varLambda_{0}}{1+\varLambda_{0}^{2}}\delta\omega+\frac{\varLambda_{0}^{2}}{1+\varLambda_{0}^{2}}h(\frac{\delta\omega}{\omega_{0}}).\label{eq: TDOS near MIT}
\end{equation}
Here, $h(\frac{\delta\omega}{\omega_{0}})=H[1-\frac{\omega^{\prime}}{\omega_{0}}]$
is the Hilbert transform of $(1-\frac{\omega^{\prime}}{\omega_{0}})$
over the range where the (leading order, linear) approximation in Eq. (\ref{eq:universal})
is valid, and it can be written as follows with
$\omega_{0}$ as the cut-off of the limited frequency range:

\begin{equation}
h(\frac{\delta\omega}{\omega_{0}})=\frac{1}{\pi}\left\{ \log\mid\frac{1+\frac{\delta\omega}{\omega_{0}}}{1-\frac{\delta\omega}{\omega_{0}}}\mid+\frac{\delta\omega}{\omega_{0}}\log\mid\frac{(\frac{\delta\omega}{\omega_{0}})^{2}-1}{(\frac{\delta\omega}{\omega_{0}})^{2}}\mid\right\} .\label{eq:h(deltaomega)}
\end{equation}
As mentioned before, this solution does not depend on the form of the disorder
distribution function, other than through the value of the parameters  $\frac{\varLambda_{0}}{1+\varLambda_{0}^{2}}$
and $\frac{\varLambda_{0}^{2}}{1+\varLambda_{0}^{2}}.$ As $\omega\rightarrow\omega_{c}$,
these quantities can be estimated simply as $\frac{\varLambda_{0}}{1+\varLambda_{0}^{2}}\sim\omega_{c}-\Delta^{\prime}(\omega_{c})$
and $\frac{\varLambda_{0}^{2}}{1+\varLambda_{0}^{2}}\sim1.$ This
condition can be satisfied for both the pseudogap and the uniform model close
to the mobility edge. Therefore, in
this limit, from the Eq.(\ref{eq: TDOS near MIT}) and Eq.(\ref{eq:h(deltaomega)})
we find%
\begin{equation}
\Delta^{\prime\prime}(\omega)\sim\varphi(\omega)\sim\left(\omega_{c}-\Delta^{\prime}(\omega_{c})+\frac{2}{\pi\omega_{0}}\right)\delta\omega-\frac{2}{\pi}\frac{\delta\omega}{\omega_{0}}\log\frac{\delta\omega}{\omega_{0}}.\label{eq:assymptotic TDOS}
\end{equation}
Remarkably, we identified logarithmic  corrections to the (linear) scaling behavior
near the Anderson metal insulator transition, obtained with TMT theory. As we show in Section V, these non analytic corrections, however, are sufficiently mild to allow for a simplified theory to be formulated by neglecting them, without sacrificing the main quantitative prediction of full TMT. 

\subsection{Analytical solution at the emergence of the pseudogap}

Here, we obtain the critical behavior of $\rho_{typ}(\omega)$ as
the pseudo-gap opens at $W=W_{c1}$. In this case the form of the disorder distribution
prohibits us using Eq. (\ref{eq:phi(omega_uniform)}) because $b(\omega_{c})=0$; we need to retain the terms to second order in $\Delta^{\prime\prime}$,  in the expansion of 
Eq. (\ref{eq:expansion 3}). Therefore, from the Eq.(\ref{eq:phi(omega_pseudogap)})
and Eq.(\ref{eq:eta}) the imaginary part of the cavity field is expressed
as

\[
\frac{\Delta^{\prime\prime}(\omega)^{2}}{(\frac{W_{c1}}{6})^{2}}=(1-\frac{1}{a(\omega)})=\delta a(\omega).
\]
Since we are interested in the behavior of the system at $W=W_{c1},$
we directly evaluate  Eq.(\ref{eq: variation of a})
as 
\begin{eqnarray*}
\delta a(\omega) & \approx & 2\left(\omega-\Delta^{\prime}(\omega)\right)\int d\epsilon P(\epsilon)\frac{1}{\omega-\Delta^{\prime}(\omega)-\epsilon}.\\
 & \approx & 2(\frac{6}{W_{c1}})^{2}\left(\omega-\Delta^{\prime}(\omega)\right)^{2}
\end{eqnarray*}
Here, $\omega-\Delta^{\prime}(\omega)$ is small near the mobility
edge $(\omega_{c}=0)$ . Therefore, the same integral equation
as Eq. (\ref{eq:universal}) can be written here in the following
form:
\begin{equation}
\Delta^{\prime\prime}(\omega)-\frac{\sqrt{2}}{\pi}\int_{-\infty}^{\infty}d\omega^{\prime}\frac{\Delta^{\prime\prime}(\omega^{\prime})}{\omega^{\prime}-\omega}=f(\omega),\label{eq:cavity fredhom2}
\end{equation}
where, $f(\omega)=\sqrt{2}\omega.$ Eq.(\ref{eq:cavity fredhom2})
has the corresponding solution which is given by
\begin{equation}
\Delta^{\prime\prime}(\omega)=\frac{\sqrt{2}}{1+2\pi^{2}}\omega+\frac{2}{1+2\pi^{2}}h(\frac{\omega}{\omega_{0}}).\label{eq:Typical critical w-1}
\end{equation}
Therefore, the critical behavior of TDOS, at critical disorder $W=W_{c1}$ where the gap opens, can be written as 
\begin{equation}
\Delta^{\prime\prime}(\omega)\sim\varphi(\omega)\sim(\frac{2}{\pi\omega_{0}}+\frac{\sqrt{2}}{1+2\pi^{2}})\omega-\frac{2}{1+2\pi^{2}}\frac{\omega}{\omega_{0}}\log\frac{\omega}{\omega_{0}}.\label{eq:assymptotic TDOS at critical point}
\end{equation}

The full analytical solution of TMT equations again provides evidence for the emergence of logarithmic
correction to scaling, even at the critical point $W=W_{c1}$. Our numerical result
in Fig.2(a) confirms that our TDOS order parameter
is assumed the same qualitative behavior at $W=W_{c1}$,  as it has been also found
near finite mobility edges with $\omega_c \neq 0$ at general $W$, 
for both the "generic" and the pseudo gap models. %
\subsection{Numerical tests of the logarithmic corrections}

Here, we show numerically that the mild logarithmic
correction can be ignored far enough critical point, 
without changing the main qualitative features of our TMT results. 
For example, the critical form of TDOS at $W=W_{c1}$ for the pseudogap model takes the form 
\[
\rho_{c}(\omega)\sim\varphi(\omega)\sim a_{1}\omega-a_{2}\omega\log\frac{\omega}{\omega_{0}}.
\]
To test this prediction, we directly plot our full numerical solution for $\rho_{typ}(\omega)/\omega$ at $W=W_{c1}$, as a function of $\log(\omega)$. The results, as shown in Fig. 5, fully support our analytical prediction for logarithmic corrections to scaling. 
\begin{figure}[H]
\includegraphics[width=3in]{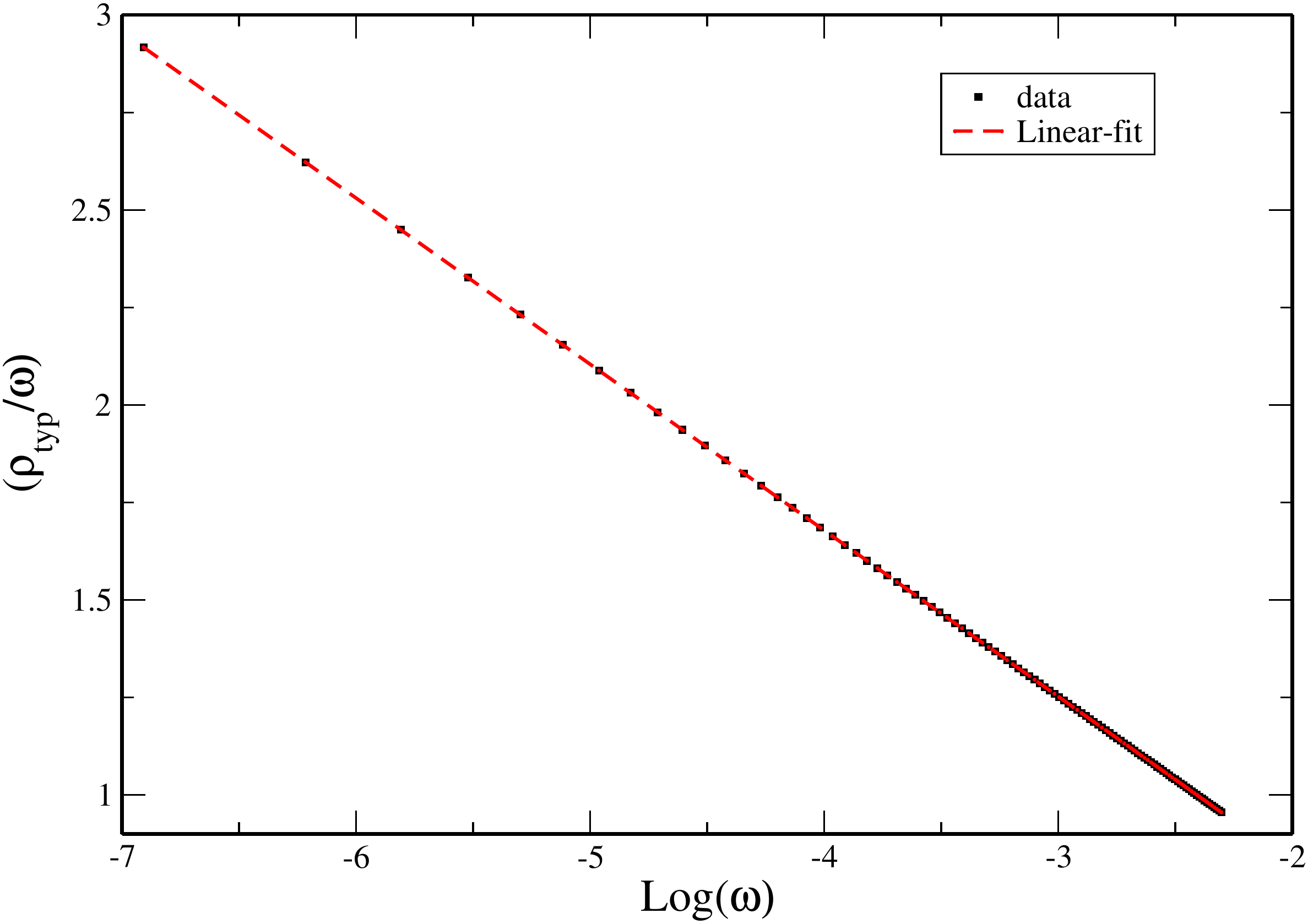}\protect\caption{The behavior of TDOS for the pseudo gap model for $W=W_c1$, on  semi-logarithmic scale, for small $0<\omega<0.1.$}
\end{figure}
Note, however, that if our result is examined only in a limited frequency interval, it can be represented as an power-law function, with an  effective exponent $\beta(\omega)$, being a weak function of frequency.

To confirm this idea, we calculate $\beta(\omega)$ both using our analytical results, and also 
using the full numerical solution of the TMT equations. From our analytical solution, we can write
\begin{equation}
\rho_{c}(\omega)\approx a_{1}\omega(1-\frac{a_{2}}{a_{1}}\log\frac{\omega}{\omega_{0}})=a_{1}\omega^{\beta(\omega)},\label{rhoc}
\end{equation}
where 
\begin{equation}
\beta(\omega)=1+\log\frac{(1-\frac{a_{2}}{a_{1}}\log\omega+\frac{a_{2}}{a_{1}}\log\omega_{0})}{\log\omega}.\label{eq:beta(omega)}
\end{equation}
As we can see from  Fig. 6, this analytical prediction is found to be in excellent quantitative comparison with the numerics. 
Within both  methods, the effective 
exponent is $\beta(\omega)$  remains close to one in the entire critical region, therefore displaying  moderate deviation
from linear behavior found if the logarithmic corrections are ignored. 
We conclude that the mild logarithmic corrections we found near mobility edges can be neglected if we are not interested in the exact values for the critical exponents. These values generally cannot be expected to be accurately predicted by a mean-field approach such as TMT.  This notion leads us to develop an  effective (simplified) Landau
theory for Anderson localization, which neglects such logarithmic corrections which preserving most qualitative trends found within TMT. 

\begin{figure}[H]
\includegraphics[width=3in]{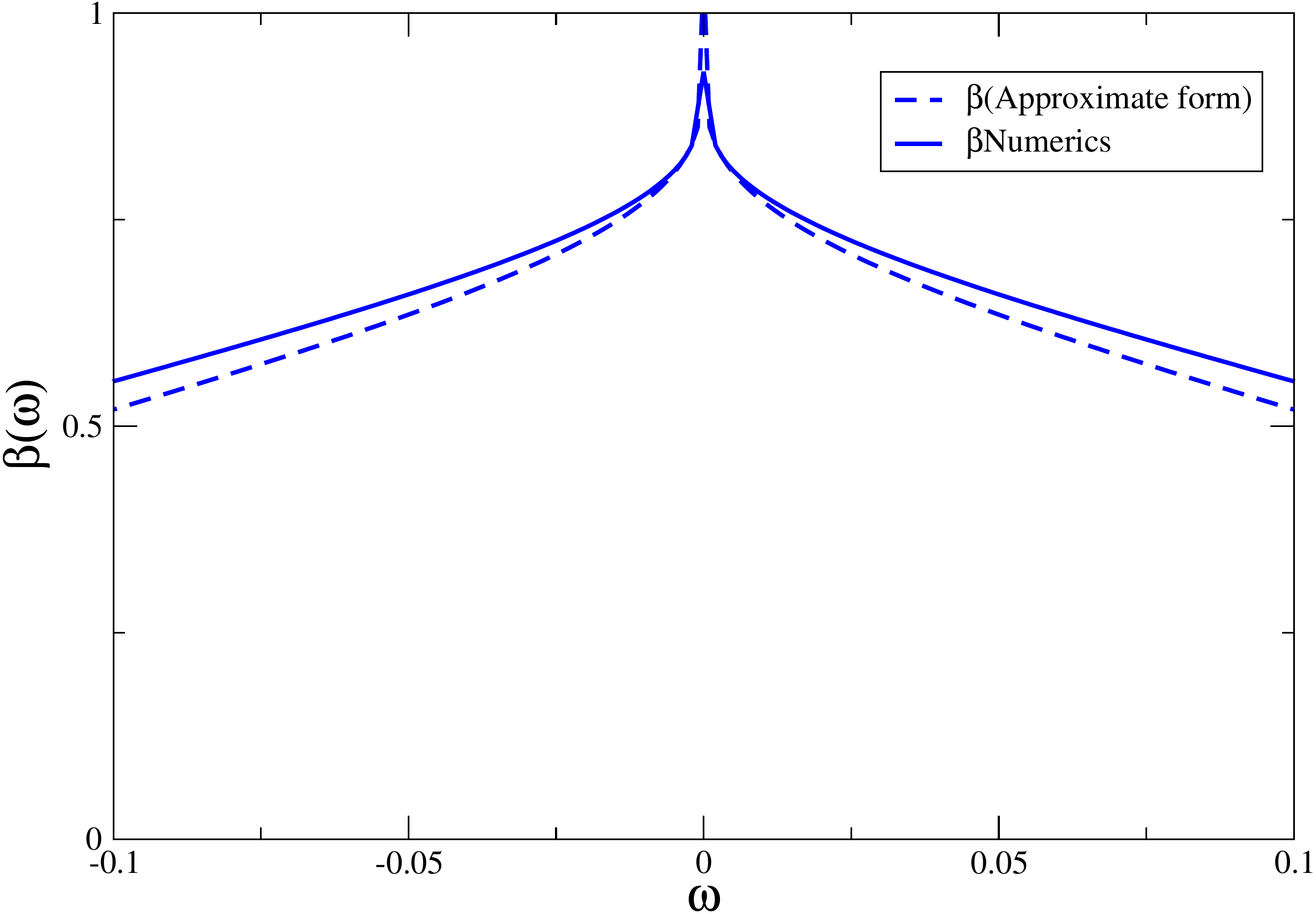}\protect\caption{The behavior of effective exponent, at the critical disorder $W=W_{c1}.$}
\end{figure}

\section{Simplified Landau theory }

In this section, we argue how the mean field solution of TMT equations
can be simplified as we are not extremely close to the critical point, where mean-field 
theories such as TMT cannot be accurate in any case. 
Since we are interested in qualitative behavior of the order parameter and other physical observables
across the phase diagram, and not only very close to the critical point, this approximation is justified and useful 
in predicting general trends. 
We have already seen that the only essential difference found within TMT, as compared to any 
mean-field theory is the emergence of mild logarithmic corrections to scaling.
Ignoring them, therefore, provides us with a simplified formulation, where the equation of state, 
i.e. the self-consistency condition for the order parameter assumes simply a polynomial form,
 as any ordinary Landau theory. In the following, we formulate such a simplified Landau theory, and show that it captures the main qualitative trends, while preserving the key difference between the two classes of models of disorder we examine. 

\subsection{Analytical prediction of the effective Landau theory}

As we have seen from Eq. (21), our TMT order-parameter satisfied a  Landau-type
equation of state, of polynomial form
\begin{equation}
r(\omega,W)+u_{1}(\omega)\varphi(\omega)+u_{2}(\omega)\varphi^{2}(\omega)+...=0.\label{eq:r(w)}
\end{equation}
To test these ideas, we use our numerical results for the TMT order parameter, and fit them to a polynomial form 
\[
\varphi(\omega)\sim\sqrt{r(\omega,W)}.
\]
These results reproduce our previous results at half-filling ($\omega =0$), where $r(0,W) \sim (W_c - W)$, as well as the general trends for the approach to mobility edges elsewhere in the phase diagram.

\subsection{Numerical fitting of the effective Landau coefficients for the pseudo-gap
model}

To test these ideas, we apply our effective Landau theory to the pseudo-gap model ($ u_1 =0$) close to the critical point. To do this, we note that according to Eq. (54), in this case to leading order
\[
r(\omega,W)\sim\rho_{typ}^{2}(\omega,W),
\]
and we can directly obtain the functional form of $r(\omega,W)$ from our numerical solution of the TMT equation. According to our Landau theory assumption, we expect it to be a smooth (analytic) function of frequency, and thus to assume a polynomial form
\[
r(\omega,W)\sim B_{0}+B_{2}\omega^{2}-B_{4}\omega^{4}+\cdots
\]
In the following, we calculate these coefficients numerically, as shown
in Fig. 7 and Fig. 8. The coefficient $B_{0}$ vanishes linearly at the transition,
consistent with previous results (See inset Fig. 4(a)). As it has been seen in Fig.
8, the coefficients $B_{2}$ and $B_{4}$ depend on $W$, but display on very weak 
dependence on the distance to the transition. As a final test, we show in Fig. 9
the behavior of TDOS, which is 
obtained both using our simplified version of TMT (simplified Landau theory)
and the exact TMT solution. The numerical results indicate
very similar behavior for order parameter within two different approaches.
These results confirm the validity of our simplified Landau theory in capturing the main trends 
obtained form the exact solution of the TMT equations. 

\begin{figure}[H]
\includegraphics[width=3in]{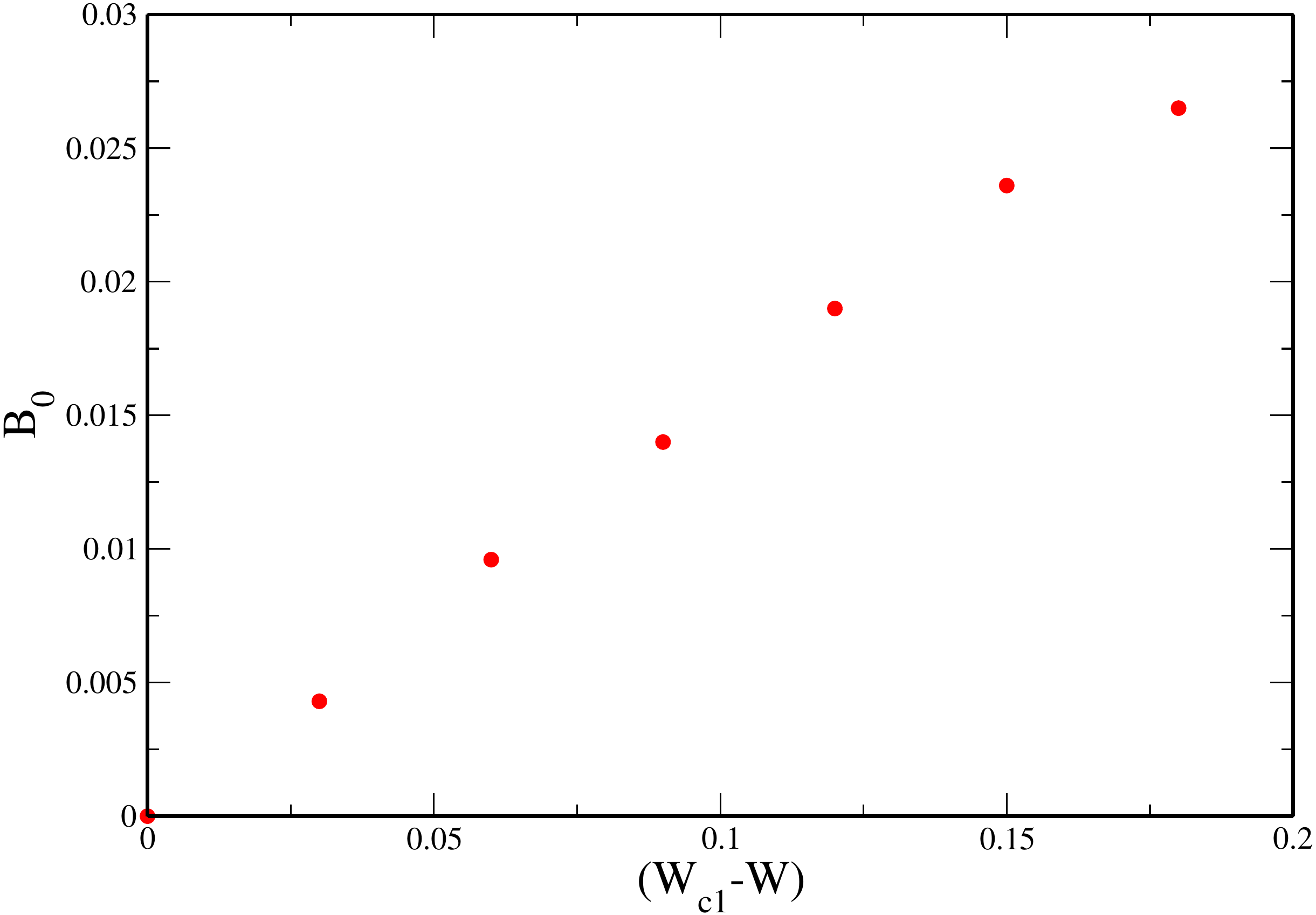}\protect\caption{The behavior of the coefficient $B_{0}$ as a function of $(W_{c1}-W)$
close to the transition. }
\end{figure}

\begin{figure}[H]
\includegraphics[width=3in]{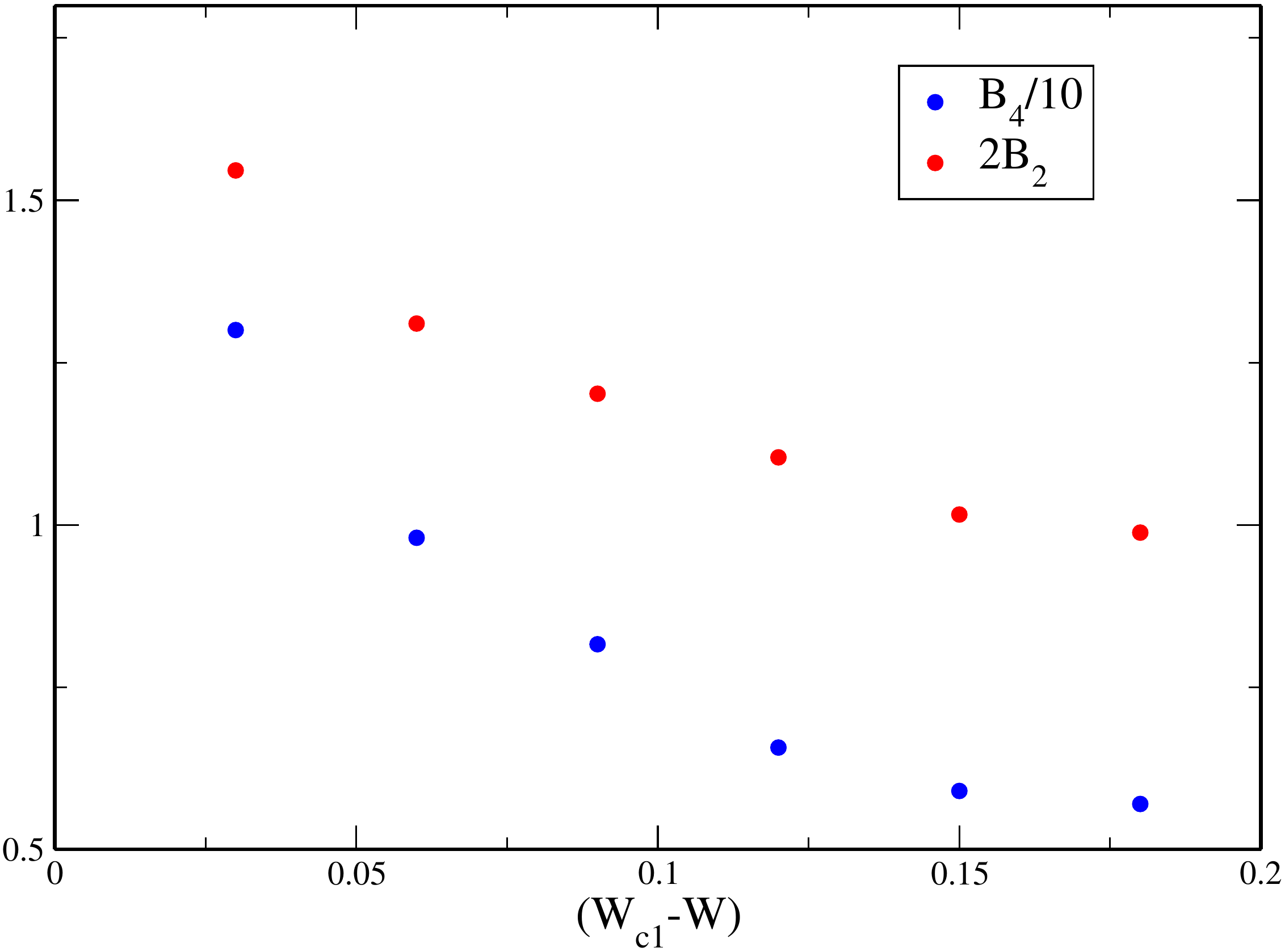}\protect\caption{The Landau coefficient $2B_{2}$ and $\frac{B_{4}}{10}$ display only weak disorder dependence 
as the transition is approached. }
\end{figure}

\begin{figure}[H]
\includegraphics[width=3in]{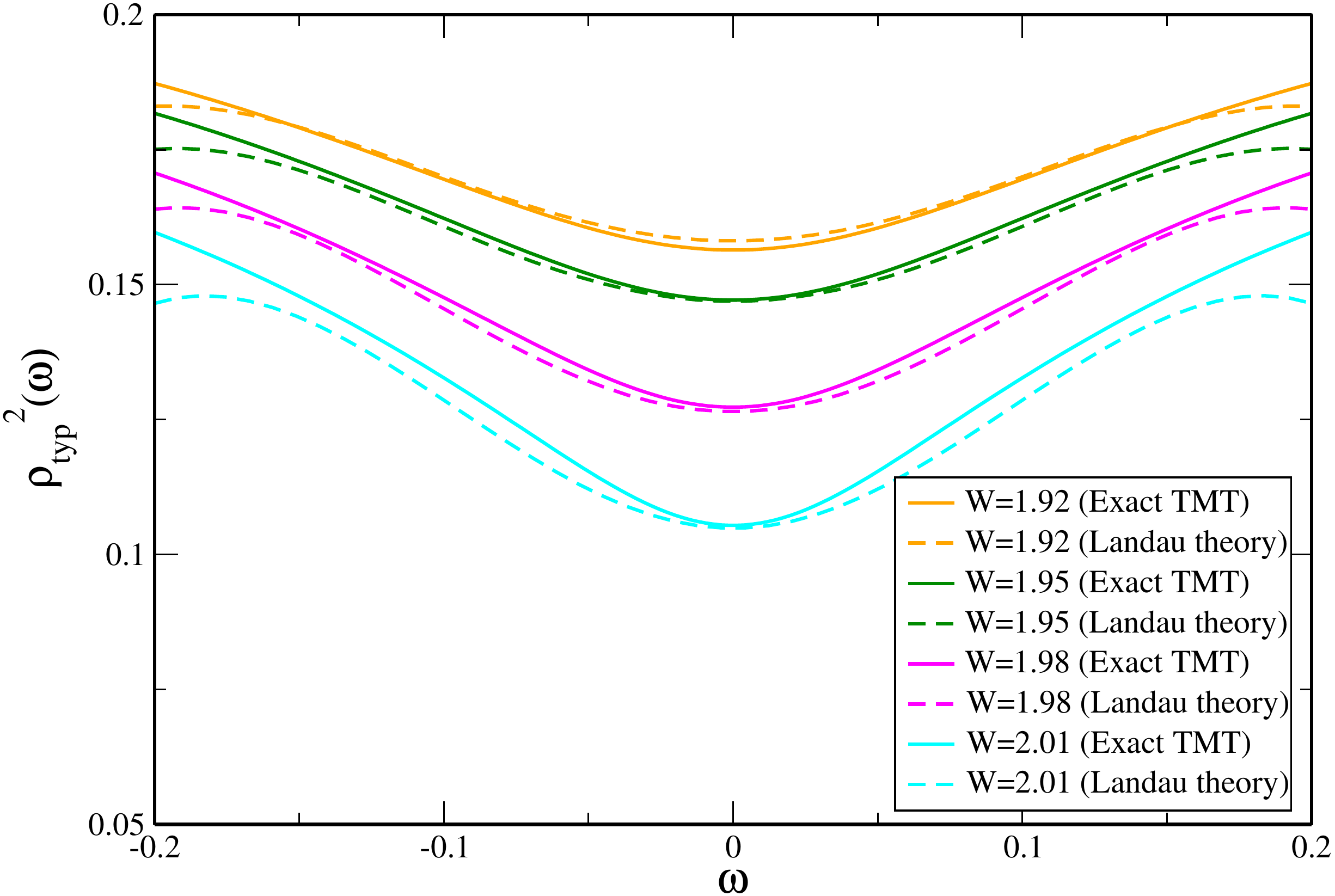}\protect\caption{Comparison between the exact numerical TMT solution and the approximate
solution, which is calculated within simplified  Landau theory. }
\end{figure}

\section{Conclusions}

In summary, we carried out a detailed TMT study of the critical behavior
for the Anderson metal-insulator transition, both analytically and numerically. Although
the exact TMT theory gives us non-analytic critical behavior, we showed
that the offending logarithmic corrections to scaling are not very
significant if we are not too close to the transition. Given the fact
that mean-field theories are generally not reliable very close
to phase transitions, our results demonstrate that for practical purposes
these subtle issues can safely be ignored, allowing us to formulate
a much-simpler Landau-like formulation for Anderson localization.
We also demonstrated that, within TMT, two different universality
classes for the critical behavior may exist, depending on the qualitative
form of disorder. We explored the opening of a soft pseudo-gap in the single-particle
density of states near the Fermi energy, which is shown to emerge
when the (renormalized) disorder is chosen to have a form appropriate for electrons
interacting through long-range Coulomb interactions. In relevant cases,
our results are found to be in excellent agreement with recent large-scale
exact diagonalization results\cite{amini2013multifractality}, as well as with recent experiments\cite{richardella2010visualizing}.
Moreover, recently developed cluster refinements of TMT demonstrated\cite{clusterTMToneandtwoD,PhysRevB.89.081107,PhysRevB.90.094208} that significant
corrections to (single site) TMT are only found very close to the Anderson transition.
All these findings provide further evidence that TMT represents
a flexible and practically useful tool for successfully describing the
main qualitative trends for physical observables, in the vicinity
of disorder-driven metal-insulator transitions.

\section*{Acknowledgments: }

The authors thanks Ali Yazdani and Stephan von Molnar for useful discussion.
This work was supported by the NSF grants DMR-1005751, DMR-1410132
and PHYS-1066293, by the National High Magnetic Field Laboratory.

\appendix

\section{Expression for the  cavity function for general band-structure models}

For simplicity, we focus on the band center ($\omega=0$), where all quantities 
we self-consistently calculate (Green's functions, cavity field, self-energies) are pure imaginary. 
In this case,  there exists a simple
relation between the typical Green's function, the self energy,  and the typical
density of states, given by the expressions
\begin{equation}
G_{typ}=-\pi\rho_{typ},\label{eq:pureimag G}
\end{equation}
and
\begin{equation}
\Sigma=-\Delta-\frac{1}{\pi\rho_{typ}}.\label{eq:self enery half filling}
\end{equation}
In order to close the self-consistent loop, we use  Eq. (\ref{eq:bare lattice Green's function}), which contains the information on the form of the electronic band-structure, through the form of the "bare" (disorder-fee) density of states
\begin{equation}
G_{typ}(\omega )=\int d\omega^{\prime}\frac{\nu_{0}(\omega^{\prime})}{\omega-\omega^{\prime}-\Sigma}.\label{eq:expansion of G}
\end{equation}
We expand the right-hand side of Eq.(\ref{eq:expansion of G}) in terms of $\frac{1}{\Sigma}$,
which remains small as we approach to transition, and write
\begin{equation}
G_{typ}=-\frac{1}{\Sigma}-(\frac{1}{\Sigma})^{3}\int d\omega^{\prime}\omega^{\prime2}\nu_{0}(\omega^{\prime})+\mathcal{O}(\frac{1}{\Sigma})^{5},\label{eq:Gtyp general}
\end{equation}
From Eq.(\ref{eq:pureimag G}) and Eq.(\ref{eq:self enery half filling}),
and keeping the leading terms in Eq.(\ref{eq:Gtyp general}), we can
obtain the general expression for cavity field as follows
\[
\Delta=C\pi\rho_{typ}+\mathcal{O}(\rho_{typ}^{2}),
\]
where
\[
C=\int d\omega^{\prime}\omega^{\prime2}\nu_{0}(\omega^{\prime}).
\]

This result shows how the coefficient $C$ can be directly calculated
at half-filling for any band-structure model. A similar a relation is valid even away from half-filling
(as we also confirmed but detailed numerical work),
but the specific numerical value  depends on the relevant non-universal parameters. 
Therefore, as in other DMFT-like theories, to capture the qualitative aspect of the critical behavior, 
it is suffices to consider the simple semi-circular model density of
states where $C=1$ for any filling and value of disorder. 

\bibliographystyle{apsrev}
\bibliography{paper}

\begin{thebibliography}{41}
\expandafter\ifx\csname natexlab\endcsname\relax\def\natexlab#1{#1}\fi
\expandafter\ifx\csname bibnamefont\endcsname\relax
  \def\bibnamefont#1{#1}\fi
\expandafter\ifx\csname bibfnamefont\endcsname\relax
  \def\bibfnamefont#1{#1}\fi
\expandafter\ifx\csname citenamefont\endcsname\relax
  \def\citenamefont#1{#1}\fi
\expandafter\ifx\csname url\endcsname\relax
  \def\url#1{\texttt{#1}}\fi
\expandafter\ifx\csname urlprefix\endcsname\relax\def\urlprefix{URL }\fi
\providecommand{\bibinfo}[2]{#2}
\providecommand{\eprint}[2][]{\url{#2}}

\bibitem[{\citenamefont{Dobrosavljevi\'c
  et~al.}(2012)\citenamefont{Dobrosavljevi\'c, Trivedi, and
  Valles~Jr}}]{dobrosavljevic2012conductor}
\bibinfo{author}{\bibfnamefont{V.}~\bibnamefont{Dobrosavljevi\'c}},
  \bibinfo{author}{\bibfnamefont{N.}~\bibnamefont{Trivedi}}, \bibnamefont{and}
  \bibinfo{author}{\bibfnamefont{J.~M.} \bibnamefont{Valles~Jr}},
  \emph{\bibinfo{title}{Conductor Insulator Quantum Phase Transitions}}
  (\bibinfo{publisher}{Oxford University Press}, \bibinfo{address}{UK},
  \bibinfo{year}{2012}).

\bibitem[{\citenamefont{Anderson}(1958)}]{anderson1958absence}
\bibinfo{author}{\bibfnamefont{P.}~\bibnamefont{Anderson}},
  \bibinfo{journal}{Physical Review} \textbf{\bibinfo{volume}{109}},
  \bibinfo{pages}{1492} (\bibinfo{year}{1958}).

\bibitem[{\citenamefont{Mott}(1990)}]{mott-book90}
\bibinfo{author}{\bibfnamefont{N.~F.} \bibnamefont{Mott}},
  \emph{\bibinfo{title}{Metal-{I}nsulator {T}ransition}}
  (\bibinfo{publisher}{Taylor \& Francis}, \bibinfo{address}{London},
  \bibinfo{year}{1990}).

\bibitem[{\citenamefont{Georges et~al.}(1996)\citenamefont{Georges, Kotliar,
  Krauth, and Rozenberg}}]{georges1996dynamical}
\bibinfo{author}{\bibfnamefont{A.}~\bibnamefont{Georges}},
  \bibinfo{author}{\bibfnamefont{G.}~\bibnamefont{Kotliar}},
  \bibinfo{author}{\bibfnamefont{W.}~\bibnamefont{Krauth}}, \bibnamefont{and}
  \bibinfo{author}{\bibfnamefont{M.}~\bibnamefont{Rozenberg}},
  \bibinfo{journal}{Rev. Mod. Phys.} \textbf{\bibinfo{volume}{68}},
  \bibinfo{pages}{13} (\bibinfo{year}{1996}).

\bibitem[{\citenamefont{Miranda and Dobrosavljevic}(2005)}]{RoP2005review}
\bibinfo{author}{\bibfnamefont{E.}~\bibnamefont{Miranda}} \bibnamefont{and}
  \bibinfo{author}{\bibfnamefont{V.}~\bibnamefont{Dobrosavljevic}},
  \bibinfo{journal}{Reports on Progress in Physics}
  \textbf{\bibinfo{volume}{68}}, \bibinfo{pages}{2337} (\bibinfo{year}{2005}).

\bibitem[{\citenamefont{Dobrosavljevi{\'c} and
  Kotliar}(1997)}]{dobrosavljevic1997mean}
\bibinfo{author}{\bibfnamefont{V.}~\bibnamefont{Dobrosavljevi{\'c}}}
  \bibnamefont{and} \bibinfo{author}{\bibfnamefont{G.}~\bibnamefont{Kotliar}},
  \bibinfo{journal}{Phys. Rev. Lett.} \textbf{\bibinfo{volume}{78}},
  \bibinfo{pages}{3943} (\bibinfo{year}{1997}).

\bibitem[{\citenamefont{Miranda and Dobrosavljevi\'{c}}(2001)}]{mirandavlad1}
\bibinfo{author}{\bibfnamefont{E.}~\bibnamefont{Miranda}} \bibnamefont{and}
  \bibinfo{author}{\bibfnamefont{V.}~\bibnamefont{Dobrosavljevi\'{c}}},
  \bibinfo{journal}{Phys. Rev. Lett.} \textbf{\bibinfo{volume}{86}},
  \bibinfo{pages}{264} (\bibinfo{year}{2001}).

\bibitem[{\citenamefont{Aguiar et~al.}(2003)\citenamefont{Aguiar, Miranda, and
  Dobrosavljevi\'{c}}}]{aguiaretal1}
\bibinfo{author}{\bibfnamefont{M.~C.~O.} \bibnamefont{Aguiar}},
  \bibinfo{author}{\bibfnamefont{E.}~\bibnamefont{Miranda}}, \bibnamefont{and}
  \bibinfo{author}{\bibfnamefont{V.}~\bibnamefont{Dobrosavljevi\'{c}}},
  \bibinfo{journal}{Phys. Rev. B} \textbf{\bibinfo{volume}{68}},
  \bibinfo{pages}{125104} (\bibinfo{year}{2003}).

\bibitem[{\citenamefont{Song et~al.}(2008)\citenamefont{Song, Wortis, and
  Atkinson}}]{atkinson2007prb}
\bibinfo{author}{\bibfnamefont{Y.}~\bibnamefont{Song}},
  \bibinfo{author}{\bibfnamefont{R.}~\bibnamefont{Wortis}}, \bibnamefont{and}
  \bibinfo{author}{\bibfnamefont{W.}~\bibnamefont{Atkinson}},
  \bibinfo{journal}{Phys. Rev. B} \textbf{\bibinfo{volume}{77}},
  \bibinfo{pages}{054202} (\bibinfo{year}{2008}).

\bibitem[{\citenamefont{Tran}(2007)}]{tran2007prb}
\bibinfo{author}{\bibfnamefont{M.-T.} \bibnamefont{Tran}},
  \bibinfo{journal}{Phys. Rev. B} \textbf{\bibinfo{volume}{76}},
  \bibinfo{pages}{245122} (\bibinfo{year}{2007}).

\bibitem[{\citenamefont{Song et~al.}(2009)\citenamefont{Song, Bulut, Wortis,
  and Atkinson}}]{atkinson2009jpc}
\bibinfo{author}{\bibfnamefont{Y.}~\bibnamefont{Song}},
  \bibinfo{author}{\bibfnamefont{S.}~\bibnamefont{Bulut}},
  \bibinfo{author}{\bibfnamefont{R.}~\bibnamefont{Wortis}}, \bibnamefont{and}
  \bibinfo{author}{\bibfnamefont{W.~A.} \bibnamefont{Atkinson}},
  \bibinfo{journal}{Journal of Physics: Condensed Matter}
  \textbf{\bibinfo{volume}{21}}, \bibinfo{pages}{385601}
  (\bibinfo{year}{2009}).

\bibitem[{\citenamefont{Andrade et~al.}(2009)\citenamefont{Andrade, Miranda,
  and Dobrosavljevic}}]{andrade09prl}
\bibinfo{author}{\bibfnamefont{E.~C.} \bibnamefont{Andrade}},
  \bibinfo{author}{\bibfnamefont{E.}~\bibnamefont{Miranda}}, \bibnamefont{and}
  \bibinfo{author}{\bibfnamefont{V.}~\bibnamefont{Dobrosavljevic}},
  \bibinfo{journal}{Phys. Rev. Lett.} \textbf{\bibinfo{volume}{102}},
  \bibinfo{pages}{206403} (\bibinfo{year}{2009}).

\bibitem[{\citenamefont{Semmler
  et~al.}(2010{\natexlab{a}})\citenamefont{Semmler, Wernsdorfer, Bissbort,
  Byczuk, and Hofstetter}}]{hofstetter2010prb-a}
\bibinfo{author}{\bibfnamefont{D.}~\bibnamefont{Semmler}},
  \bibinfo{author}{\bibfnamefont{J.}~\bibnamefont{Wernsdorfer}},
  \bibinfo{author}{\bibfnamefont{U.}~\bibnamefont{Bissbort}},
  \bibinfo{author}{\bibfnamefont{K.}~\bibnamefont{Byczuk}}, \bibnamefont{and}
  \bibinfo{author}{\bibfnamefont{W.}~\bibnamefont{Hofstetter}},
  \bibinfo{journal}{Phys. Rev. B} \textbf{\bibinfo{volume}{82}},
  \bibinfo{pages}{235115} (\bibinfo{year}{2010}{\natexlab{a}}).

\bibitem[{\citenamefont{Semmler
  et~al.}(2010{\natexlab{b}})\citenamefont{Semmler, Byczuk, and
  Hofstetter}}]{hofstetter2010prb-b}
\bibinfo{author}{\bibfnamefont{D.}~\bibnamefont{Semmler}},
  \bibinfo{author}{\bibfnamefont{K.}~\bibnamefont{Byczuk}}, \bibnamefont{and}
  \bibinfo{author}{\bibfnamefont{W.}~\bibnamefont{Hofstetter}},
  \bibinfo{journal}{Phys. Rev. B} \textbf{\bibinfo{volume}{81}},
  \bibinfo{pages}{115111} (\bibinfo{year}{2010}{\natexlab{b}}).

\bibitem[{\citenamefont{Semmler et~al.}(2011)\citenamefont{Semmler, Byczuk, and
  Hofstetter}}]{hofstetter2011prb}
\bibinfo{author}{\bibfnamefont{D.}~\bibnamefont{Semmler}},
  \bibinfo{author}{\bibfnamefont{K.}~\bibnamefont{Byczuk}}, \bibnamefont{and}
  \bibinfo{author}{\bibfnamefont{W.}~\bibnamefont{Hofstetter}},
  \bibinfo{journal}{Phys. Rev. B} \textbf{\bibinfo{volume}{84}},
  \bibinfo{pages}{115113} (\bibinfo{year}{2011}).

\bibitem[{\citenamefont{Aguiar and Dobrosavljevi\ifmmode~\acute{c}\else
  \'{c}\fi{}}(2013)}]{aguiar2013prl}
\bibinfo{author}{\bibfnamefont{M.~C.~O.} \bibnamefont{Aguiar}}
  \bibnamefont{and}
  \bibinfo{author}{\bibfnamefont{V.}~\bibnamefont{Dobrosavljevi\ifmmode~\acute{c}\else
  \'{c}\fi{}}}, \bibinfo{journal}{Phys. Rev. Lett.}
  \textbf{\bibinfo{volume}{110}}, \bibinfo{pages}{066401}
  (\bibinfo{year}{2013}).

\bibitem[{\citenamefont{Potthoff and Nolting}(1999)}]{potthoff1999prb}
\bibinfo{author}{\bibfnamefont{M.}~\bibnamefont{Potthoff}} \bibnamefont{and}
  \bibinfo{author}{\bibfnamefont{W.}~\bibnamefont{Nolting}},
  \bibinfo{journal}{Phys. Rev. B} \textbf{\bibinfo{volume}{59}},
  \bibinfo{pages}{2549} (\bibinfo{year}{1999}).

\bibitem[{\citenamefont{Helmes et~al.}(2008)\citenamefont{Helmes, Costi, and
  Rosch}}]{rosch2008prl}
\bibinfo{author}{\bibfnamefont{R.}~\bibnamefont{Helmes}},
  \bibinfo{author}{\bibfnamefont{T.}~\bibnamefont{Costi}}, \bibnamefont{and}
  \bibinfo{author}{\bibfnamefont{A.}~\bibnamefont{Rosch}},
  \bibinfo{journal}{Phys. Rev. Lett.} \textbf{\bibinfo{volume}{101}},
  \bibinfo{pages}{066802} (\bibinfo{year}{2008}).

\bibitem[{\citenamefont{Snoek et~al.}(2008)\citenamefont{Snoek, Titvinidze,
  Toke, Byczuk, and Hofstetter}}]{hofstetter2008njp}
\bibinfo{author}{\bibfnamefont{M.}~\bibnamefont{Snoek}},
  \bibinfo{author}{\bibfnamefont{I.}~\bibnamefont{Titvinidze}},
  \bibinfo{author}{\bibfnamefont{C.}~\bibnamefont{Toke}},
  \bibinfo{author}{\bibfnamefont{K.}~\bibnamefont{Byczuk}}, \bibnamefont{and}
  \bibinfo{author}{\bibfnamefont{W.}~\bibnamefont{Hofstetter}},
  \bibinfo{journal}{New Journal of Physics} \textbf{\bibinfo{volume}{10}},
  \bibinfo{pages}{093008} (\bibinfo{year}{2008}).

\bibitem[{\citenamefont{Thouless et~al.}(1977)\citenamefont{Thouless, Anderson,
  and Palmer}}]{TAP}
\bibinfo{author}{\bibfnamefont{D.~J.} \bibnamefont{Thouless}},
  \bibinfo{author}{\bibfnamefont{P.~W.} \bibnamefont{Anderson}},
  \bibnamefont{and} \bibinfo{author}{\bibfnamefont{R.~G.}
  \bibnamefont{Palmer}}, \bibinfo{journal}{Philosophical Magazine}
  \textbf{\bibinfo{volume}{35}}, \bibinfo{pages}{137} (\bibinfo{year}{1977}).

\bibitem[{\citenamefont{M.Janssen}(1998)}]{re:Janssen98}
\bibinfo{author}{\bibnamefont{M.Janssen}}, \bibinfo{journal}{Phys. Rep.}
  \textbf{\bibinfo{volume}{295}}, \bibinfo{pages}{1} (\bibinfo{year}{1998}).

\bibitem[{\citenamefont{Dobrosavljevi{\'c}
  et~al.}(2003)\citenamefont{Dobrosavljevi{\'c}, Pastor, and
  Nikolic}}]{pastor2001tmt}
\bibinfo{author}{\bibfnamefont{V.}~\bibnamefont{Dobrosavljevi{\'c}}},
  \bibinfo{author}{\bibfnamefont{A.~A.} \bibnamefont{Pastor}},
  \bibnamefont{and} \bibinfo{author}{\bibfnamefont{B.~K.}
  \bibnamefont{Nikolic}}, \bibinfo{journal}{Europh. Lett.}
  \textbf{\bibinfo{volume}{62}}, \bibinfo{pages}{76} (\bibinfo{year}{2003}).

\bibitem[{\citenamefont{Byczuk et~al.}(2005)\citenamefont{Byczuk, Hofstetter,
  and Vollhardt}}]{hofstetter2005prl}
\bibinfo{author}{\bibfnamefont{K.}~\bibnamefont{Byczuk}},
  \bibinfo{author}{\bibfnamefont{W.}~\bibnamefont{Hofstetter}},
  \bibnamefont{and}
  \bibinfo{author}{\bibfnamefont{D.}~\bibnamefont{Vollhardt}},
  \bibinfo{journal}{Phys. Rev. Lett.} \textbf{\bibinfo{volume}{94}},
  \bibinfo{pages}{056404} (\bibinfo{year}{2005}).

\bibitem[{\citenamefont{Aguiar et~al.}(2006)\citenamefont{Aguiar,
  Dobrosavljevi\ifmmode~\acute{c}\else \'{c}\fi{}, Abrahams, and
  Kotliar}}]{aguiar2006prb}
\bibinfo{author}{\bibfnamefont{M.}~\bibnamefont{Aguiar}},
  \bibinfo{author}{\bibfnamefont{V.}~\bibnamefont{Dobrosavljevi\ifmmode~\acute{c}\else
  \'{c}\fi{}}}, \bibinfo{author}{\bibfnamefont{E.}~\bibnamefont{Abrahams}},
  \bibnamefont{and} \bibinfo{author}{\bibfnamefont{G.}~\bibnamefont{Kotliar}},
  \bibinfo{journal}{Phys. Rev. B} \textbf{\bibinfo{volume}{73}},
  \bibinfo{pages}{115117} (\bibinfo{year}{2006}).

\bibitem[{\citenamefont{Byczuk et~al.}(2009)\citenamefont{Byczuk, Hofstetter,
  and Vollhardt}}]{hofstetter2009prl}
\bibinfo{author}{\bibfnamefont{K.}~\bibnamefont{Byczuk}},
  \bibinfo{author}{\bibfnamefont{W.}~\bibnamefont{Hofstetter}},
  \bibnamefont{and}
  \bibinfo{author}{\bibfnamefont{D.}~\bibnamefont{Vollhardt}},
  \bibinfo{journal}{Phys. Rev. Lett.} \textbf{\bibinfo{volume}{102}},
  \bibinfo{pages}{146403} (\bibinfo{year}{2009}).

\bibitem[{\citenamefont{Aguiar et~al.}(2009)\citenamefont{Aguiar,
  Dobrosavljevi\ifmmode~\acute{c}\else \'{c}\fi{}, Abrahams, and
  Kotliar}}]{aguiar2009prl}
\bibinfo{author}{\bibfnamefont{M.}~\bibnamefont{Aguiar}},
  \bibinfo{author}{\bibfnamefont{V.}~\bibnamefont{Dobrosavljevi\ifmmode~\acute{c}\else
  \'{c}\fi{}}}, \bibinfo{author}{\bibfnamefont{E.}~\bibnamefont{Abrahams}},
  \bibnamefont{and} \bibinfo{author}{\bibfnamefont{G.}~\bibnamefont{Kotliar}},
  \bibinfo{journal}{Phys. Rev. Lett.} \textbf{\bibinfo{volume}{102}},
  \bibinfo{pages}{156402} (\bibinfo{year}{2009}).

\bibitem[{\citenamefont{Oliveira et~al.}(2014)\citenamefont{Oliveira, Aguiar,
  and Dobrosavljevi\ifmmode~\acute{c}\else \'{c}\fi{}}}]{aguiar2014prb}
\bibinfo{author}{\bibfnamefont{W.~S.} \bibnamefont{Oliveira}},
  \bibinfo{author}{\bibfnamefont{M.~C.~O.} \bibnamefont{Aguiar}},
  \bibnamefont{and}
  \bibinfo{author}{\bibfnamefont{V.}~\bibnamefont{Dobrosavljevi\ifmmode~\acute{c}\else
  \'{c}\fi{}}}, \bibinfo{journal}{Phys. Rev. B} \textbf{\bibinfo{volume}{89}},
  \bibinfo{pages}{165138} (\bibinfo{year}{2014}).

\bibitem[{\citenamefont{Dobrosavljevic}(2010)}]{dobrosavljevic2010typical}
\bibinfo{author}{\bibfnamefont{V.}~\bibnamefont{Dobrosavljevic}},
  \bibinfo{journal}{Int. J. Mod. Phys. B} \textbf{\bibinfo{volume}{24}},
  \bibinfo{pages}{1680} (\bibinfo{year}{2010}).

\bibitem[{\citenamefont{Richardella et~al.}(2010)\citenamefont{Richardella,
  Roushan, Mack, Zhou, Huse, Awschalom, and
  Yazdani}}]{richardella2010visualizing}
\bibinfo{author}{\bibfnamefont{A.}~\bibnamefont{Richardella}},
  \bibinfo{author}{\bibfnamefont{P.}~\bibnamefont{Roushan}},
  \bibinfo{author}{\bibfnamefont{S.}~\bibnamefont{Mack}},
  \bibinfo{author}{\bibfnamefont{B.}~\bibnamefont{Zhou}},
  \bibinfo{author}{\bibfnamefont{D.}~\bibnamefont{Huse}},
  \bibinfo{author}{\bibfnamefont{D.}~\bibnamefont{Awschalom}},
  \bibnamefont{and} \bibinfo{author}{\bibfnamefont{A.}~\bibnamefont{Yazdani}},
  \bibinfo{journal}{Science} \textbf{\bibinfo{volume}{327}},
  \bibinfo{pages}{665} (\bibinfo{year}{2010}).

\bibitem[{\citenamefont{Efros and Shklovskii}(1975)}]{efros1975coulomb}
\bibinfo{author}{\bibfnamefont{A.}~\bibnamefont{Efros}} \bibnamefont{and}
  \bibinfo{author}{\bibfnamefont{B.}~\bibnamefont{Shklovskii}},
  \bibinfo{journal}{Journal of Physics C: Solid State Physics}
  \textbf{\bibinfo{volume}{8}}, \bibinfo{pages}{L49} (\bibinfo{year}{1975}).

\bibitem[{\citenamefont{Efros}(1976)}]{efros1976coulomb}
\bibinfo{author}{\bibfnamefont{A.}~\bibnamefont{Efros}},
  \bibinfo{journal}{Journal of Physics C: Solid State Physics}
  \textbf{\bibinfo{volume}{9}}, \bibinfo{pages}{2021} (\bibinfo{year}{1976}).

\bibitem[{\citenamefont{Efros and Pollak}(1985)}]{efros1985electron}
\bibinfo{author}{\bibfnamefont{A.~L.} \bibnamefont{Efros}} \bibnamefont{and}
  \bibinfo{author}{\bibfnamefont{M.}~\bibnamefont{Pollak}},
  \emph{\bibinfo{title}{Electron-electron interactions in disordered systems}}
  (\bibinfo{publisher}{North Holland}, \bibinfo{year}{1985}).

\bibitem[{\citenamefont{Economou}(2006)}]{Economou2006}
\bibinfo{author}{\bibfnamefont{E.~N.} \bibnamefont{Economou}},
  \emph{\bibinfo{title}{Green\'s Functions in Quantum Physics}}
  (\bibinfo{publisher}{Springer, Berlin}, \bibinfo{year}{2006}).

\bibitem[{\citenamefont{Evers and Mirlin}(2008)}]{evers2008anderson}
\bibinfo{author}{\bibfnamefont{F.}~\bibnamefont{Evers}} \bibnamefont{and}
  \bibinfo{author}{\bibfnamefont{A.}~\bibnamefont{Mirlin}},
  \bibinfo{journal}{Reviews of Modern Physics} \textbf{\bibinfo{volume}{80}},
  \bibinfo{pages}{1355} (\bibinfo{year}{2008}).

\bibitem[{\citenamefont{Massey and Lee}(1996)}]{massey1996prl}
\bibinfo{author}{\bibfnamefont{J.}~\bibnamefont{Massey}} \bibnamefont{and}
  \bibinfo{author}{\bibfnamefont{M.}~\bibnamefont{Lee}},
  \bibinfo{journal}{Phys. Rev. Lett.} \textbf{\bibinfo{volume}{77}},
  \bibinfo{pages}{3399} (\bibinfo{year}{1996}).

\bibitem[{\citenamefont{Pramudya et~al.}(2011)\citenamefont{Pramudya,
  Terletska, Pankov, Manousakis, and Dobrosavljevi\ifmmode~\acute{c}\else
  \'{c}\fi{}}}]{2011prb}
\bibinfo{author}{\bibfnamefont{Y.}~\bibnamefont{Pramudya}},
  \bibinfo{author}{\bibfnamefont{H.}~\bibnamefont{Terletska}},
  \bibinfo{author}{\bibfnamefont{S.}~\bibnamefont{Pankov}},
  \bibinfo{author}{\bibfnamefont{E.}~\bibnamefont{Manousakis}},
  \bibnamefont{and}
  \bibinfo{author}{\bibfnamefont{V.}~\bibnamefont{Dobrosavljevi\ifmmode~\acute{c}\else
  \'{c}\fi{}}}, \bibinfo{journal}{Phys. Rev. B} \textbf{\bibinfo{volume}{84}},
  \bibinfo{pages}{125120} (\bibinfo{year}{2011}).

\bibitem[{\citenamefont{Amini et~al.}(2014)\citenamefont{Amini, Kravtsov, and
  M$\ddot{u}$ller}}]{amini2013multifractality}
\bibinfo{author}{\bibfnamefont{M.}~\bibnamefont{Amini}},
  \bibinfo{author}{\bibfnamefont{V.~E.} \bibnamefont{Kravtsov}},
  \bibnamefont{and}
  \bibinfo{author}{\bibfnamefont{M.}~\bibnamefont{M$\ddot{u}$ller}},
  \bibinfo{journal}{New Journal of Physics} \textbf{\bibinfo{volume}{16}},
  \bibinfo{pages}{015022} (\bibinfo{year}{2014}).

\bibitem[{\citenamefont{Polyanin and Manzhirov}(2008)}]{polyanin2008handbook}
\bibinfo{author}{\bibfnamefont{A.~D.} \bibnamefont{Polyanin}} \bibnamefont{and}
  \bibinfo{author}{\bibfnamefont{A.~V.} \bibnamefont{Manzhirov}},
  \emph{\bibinfo{title}{Handbook of integral equations}}
  (\bibinfo{publisher}{Chapman and Hall/CRC}, \bibinfo{year}{2008}).

\bibitem[{\citenamefont{Ekuma et~al.}(2014{\natexlab{a}})\citenamefont{Ekuma,
  Terletska, Meng, Moreno, Jarrell, Mahmoudian, and
  Dobrosavljevic}}]{clusterTMToneandtwoD}
\bibinfo{author}{\bibfnamefont{C.~E.} \bibnamefont{Ekuma}},
  \bibinfo{author}{\bibfnamefont{H.}~\bibnamefont{Terletska}},
  \bibinfo{author}{\bibfnamefont{Z.~Y.} \bibnamefont{Meng}},
  \bibinfo{author}{\bibfnamefont{J.}~\bibnamefont{Moreno}},
  \bibinfo{author}{\bibfnamefont{M.}~\bibnamefont{Jarrell}},
  \bibinfo{author}{\bibfnamefont{S.}~\bibnamefont{Mahmoudian}},
  \bibnamefont{and}
  \bibinfo{author}{\bibfnamefont{V.}~\bibnamefont{Dobrosavljevic}},
  \bibinfo{journal}{Journal of Physics: Condensed Matter}
  \textbf{\bibinfo{volume}{26}}, \bibinfo{pages}{274209}
  (\bibinfo{year}{2014}{\natexlab{a}}).

\bibitem[{\citenamefont{Ekuma et~al.}(2014{\natexlab{b}})\citenamefont{Ekuma,
  Terletska, Tam, Meng, Moreno, and Jarrell}}]{PhysRevB.89.081107}
\bibinfo{author}{\bibfnamefont{C.~E.} \bibnamefont{Ekuma}},
  \bibinfo{author}{\bibfnamefont{H.}~\bibnamefont{Terletska}},
  \bibinfo{author}{\bibfnamefont{K.-M.} \bibnamefont{Tam}},
  \bibinfo{author}{\bibfnamefont{Z.-Y.} \bibnamefont{Meng}},
  \bibinfo{author}{\bibfnamefont{J.}~\bibnamefont{Moreno}}, \bibnamefont{and}
  \bibinfo{author}{\bibfnamefont{M.}~\bibnamefont{Jarrell}},
  \bibinfo{journal}{Phys. Rev. B} \textbf{\bibinfo{volume}{89}},
  \bibinfo{pages}{081107} (\bibinfo{year}{2014}{\natexlab{b}}).

\bibitem[{\citenamefont{Terletska et~al.}(2014)\citenamefont{Terletska, Ekuma,
  Moore, Tam, Moreno, and Jarrell}}]{PhysRevB.90.094208}
\bibinfo{author}{\bibfnamefont{H.}~\bibnamefont{Terletska}},
  \bibinfo{author}{\bibfnamefont{C.~E.} \bibnamefont{Ekuma}},
  \bibinfo{author}{\bibfnamefont{C.}~\bibnamefont{Moore}},
  \bibinfo{author}{\bibfnamefont{K.-M.} \bibnamefont{Tam}},
  \bibinfo{author}{\bibfnamefont{J.}~\bibnamefont{Moreno}}, \bibnamefont{and}
  \bibinfo{author}{\bibfnamefont{M.}~\bibnamefont{Jarrell}},
  \bibinfo{journal}{Phys. Rev. B} \textbf{\bibinfo{volume}{90}},
  \bibinfo{pages}{094208} (\bibinfo{year}{2014}).

\end{thebibliography}

\end{document}